\documentclass[11pt,a4paper]{amsart}
\pdfoutput=1
\usepackage[english]{babel}
\usepackage[utf8]{inputenc}
\usepackage{amssymb,amsmath}
\usepackage{mathtools}
\usepackage{graphicx}
\usepackage{bm,bbm}
\usepackage{color}
\usepackage{tikz}
\usepackage{ifthen}
\usetikzlibrary{snakes}
\usetikzlibrary{patterns}
\usepackage{algorithm}
\usepackage{setspace}
\usepackage[noend]{algpseudocode}
\usepackage{pgfplots}
\usepackage{enumitem}
\usepackage{array,blkarray}
\usepackage{bigdelim}
\usepackage{listings} 
\usepackage[hidelinks]{hyperref}
\usepackage{amssymb}
\usepackage{isotope}
\usepackage{stmaryrd}
\usepackage[]{placeins}
\usepackage[normalem]{ulem}
\usepackage[numbers]{natbib}
\usepackage{tikz}
\usepackage{pgfplots}
\usepackage{pgfplotstable}
\usepgfplotslibrary{groupplots}
\usetikzlibrary{shapes,arrows.meta}
\usetikzlibrary{external}
\tikzset{external/system call={lualatex -shell-escape -halt-on-error -interaction=batchmode -jobname "\image" "\texsource"}}
\usepackage{ifthen} 
\newboolean{professormode}
\setboolean{professormode}{true}
\usepackage{caption}
\usepackage{subcaption}
\DeclareMathAlphabet{\mathbbold}{U}{bbold}{m}{n}
\newcommand{\pluseq}{\mathrel{+}=}

\algnewcommand{\LineComment}[1]{\State \(\triangleright\) #1}
\newcolumntype{L}[1]{>{\raggedright\let\newline\\\arraybackslash\hspace{0pt}}m{#1}}
\newcolumntype{C}[1]{>{\centering\let\newline\\\arraybackslash\hspace{0pt}}m{#1}}
\newcolumntype{R}[1]{>{\raggedleft\let\newline\\\arraybackslash\hspace{0pt}}m{#1}}
\newtheorem{theorem}{Theorem}

\theoremstyle{definition}

\theoremstyle{remark}
\newtheorem{remark}[theorem]{Remark}

\numberwithin{theorem}{section}
\numberwithin{equation}{section}
\numberwithin{table}{section}
\numberwithin{figure}{section}

\DeclareMathOperator{\rank}{rank}
\definecolor{myBlue}{RGB}{113,104,238} 
\definecolor{myGreen}{RGB}{154,205,50} 
\definecolor{myGreen2}{RGB}{114,175,30} 
\definecolor{myRed}{RGB}{180,50,50}  
\definecolor{myOrange}{RGB}{225,92,22} 
\definecolor{lgray}{RGB}{200,200,200} 
\definecolor{llgray}{RGB}{155,155,155} 
\definecolor{mycolor1}{rgb}{0.00000,0.44700,0.74100}%
\definecolor{mycolor2}{rgb}{0.85000,0.32500,0.09800}%
\definecolor{mycolor3}{rgb}{0.92900,0.69400,0.12500}%
\definecolor{mycolor4}{rgb}{0.49400,0.18400,0.55600}%
\definecolor{mycolor5}{rgb}{0.46600,0.67400,0.18800}%
\definecolor{mycolor6}{rgb}{0.30100,0.74500,0.93300}%
\definecolor{mycolor7}{rgb}{0.63500,0.07800,0.18400}%
\newcommand\N{\mathbb N}

\newcommand\R{\mathbb R}
\newcommand\Z{\mathbb Z}

\DeclareMathOperator{\diag}{diag}

\newcommand{\x}{\mathbf{x}}

\begin{document}
\title[Rank-$1$ convexification of incremental damage at finite strains]{Multidimensional rank-one convexification of incremental damage models at finite strains}

\author[]{D.~Balzani$^{*}$, M.~Köhler$^{*}$, T.~Neumeier$^{\dagger}$, M.~A.~Peter$^{\ddagger}$,  D.~Peterseim$^{\ddagger}$}
\address{${}^{*}$ Chair of Continuum Mechanics, Ruhr-Universit\"at Bochum, Universit\"atsstr.~150, 44801~Bochum, Germany}
\email{\{daniel.balzani, maximilian.koehler\}@rub.de,}
\address{${}^{\dagger}$ Institute of Mathematics, University of Augsburg, Universit\"atsstr.~12a, 86159 Augsburg, Germany}
\email{timo.neumeier@uni-a.de}
\address{${}^{\ddagger}$ Institute of Mathematics \& Centre for Advanced Analytics and Predictive Sciences (CAAPS), University of Augsburg, Universit\"atsstr.~12a, 86159 Augsburg, Germany}
\email{\{malte.peter, daniel.peterseim\}@uni-a.de}
\thanks{\textit{Acknowledgments.} The authors acknowledge funding by the Deutsche Forschungsgemeinschaft (DFG) within the Priority Program 2256 (“Variational Methods for Predicting Complex Phenomena in Engineering Structures and Materials”), Project ID 441154176, reference IDs BA2823/17-1, PE1464/7-1, and PE2143/5-1. Furthermore, the authors acknowledge the free and open source community of the Julia programming language, especially of Ferrite.jl. }

\date{\today}
\keywords{Numerical relaxation, Rank-One convexification, Continuum damage mechanics}

\begin{abstract}
This paper presents computationally feasible rank-one relaxation algorithms for the efficient simulation of a time-incremental damage model with nonconvex incremental stress potentials in multiple spatial dimensions.
While the standard model suffers from numerical issues due to the lack of convexity, the relaxation by rank-one convexification prevents non-existence of minimizers and mesh dependence of the solutions of finite element discretizations. 
By the combination, modification and parallelization of the underlying convexification algorithms, the novel approach becomes computationally feasible. 
A descent method and a Newton scheme enhanced by step-size control prevent stability issues related to local minima in the energy landscape and the computation of derivatives. 
Numerical techniques for the construction of continuous derivatives of the approximated rank-one convex envelope are discussed.
A series of numerical experiments demonstrates the ability of the computationally relaxed model to capture softening effects and the mesh independence of the computed approximations. 
An interpretation in terms of microstructural damage evolution is given, based on the rank-one lamination process.
\end{abstract}

\maketitle
%
{\tiny {\bf Key words.} Numerical relaxation, Rank-One convexification, Continuum damage mechanics 
}\\
\indent
{\tiny {\bf AMS subject classifications.} {\bf 65K10}, {\bf74G65}, {\bf 74A45}} 
%
%
%

%

\section{Introduction}\label{sec:Introduction}
The degradation of materials is a critical issue in continuum mechanics.
Hence, the modeling of damage processes is important.
The formation and evolution of voids in a material on the microscale deteriorates the mechanical properties and thus weakens the material on the macroscale.
The microscopic damage affects the macroscale material response and effects such as stress softening and strain softening occur.
The first one describes a reduction of material stiffness with an increase of stress, while strain softening refers to the reduction of stress with increasing strain.
The phenomenological scalar-valued \((1-D)\)-approach for modeling damage and capturing these softening effects was introduced in \cite{Kac:1958:rtc}.
This approach has been rigorously analyzed and further extended to the finite strain setting in \cite{Sim:1987:ftf,Mie:1995:dcd}.
While this model describes real materials accurately, the direct description of strain softening possesses mathematical complications.
Namely, strain softening often implies some degree of non-convexity in the underlying incremental stress potential (generalized energy density) and, thus, mesh dependence will be observed.
Mesh dependence refers to the behaviour that integral values of the finite element simulation will change despite choosing a very fine mesh, see e.g.~\cite[Section 6.4.2]{deCriRemVer::nfe}, \cite[Section 11.2]{Mur:2012:cdm}.
Another mathematical perspective on this issue is that the underlying Euler--Lagrange equation looses ellipticity.
To analyze this, a framework that yields a pseudo-elastic potential is required, since most mathematical results on the existence of minimizers cover the case of a potential formulation with dependence on the gradient.
The appropriate tool is the incremental variational framework, which was introduced in a series of papers, see e.g.~\cite{Hac:1997:gsm,OrtRep:1999:nem,OrtSta:1999:vfv,CarHacMie:2002:ncp,MieTheLev:2002:vfr}.
This framework provides thermodynamically consistent potentials and \textit{condenses} the dependence on the dissipative internal variable out of the problem, leading to a pseudo-elastic formulation per incremental step.
To circumvent the loss of ellipticity and, thus, mesh dependence, various regularization methodologies exist.
The temporal regularization introduces viscous terms, see e.g.~\cite{FarOliCer:1998:spv,SufLubCom:2003:dmd,LanKurMos:2022:hrc,Nee:1988:mrd,LanJunMos:2018:qdm}, while spatial regularization introduces non-locality either through neighbourhood integrals or gradient extensions regarding the damage variable \cite{PeeGeeBorBre:2001:ccn,KieWafSprMen:2018:gdm,DimHac:2008:mge,DimHac:2011:rfd,JunRieBal:2022:ern, RieBal:2022:sel}.
In the context of crack propagation, an alternative is the incorporation of non-local normalizations of crack surfaces, see \cite{PanOrt:2012:eab,WinBal:2022:scp}.
All these methodologies have in common that an additional parameter is introduced, which corresponds either to viscosity or an internal length scale.

The identification of these parameters is a difficult task and sometimes contradictory to the required numerical values.
Another regularization technique is relaxation, which is the focus of this paper.
In relaxation, the original non-convex incremental stress potential is replaced by a (semi)convex envelope, which is the greatest (semi)convex function below the original function.
This was done for the one dimensional setting of damage at finite strains in \cite{BalOrt:2012:riv, SchBal:2016:riv, KohNeuMelPetPetBal:2022:acm, KohBal:2023:emr}, where all semiconvex notions coincide with convexity.
For higher spatial dimensions, the microsphere approach, cf.~\cite{Baz:1984:mms,MieGokLul:2004:mar,FreIhl:2010:gom}, was used to obtain a three-dimensional material model via relaxed one-dimensional contributions.
However, in the finite-strain setting of continuum damage mechanics, a full higher dimensional relaxation with a suitable semiconvex envelope has not been accomplished so far.
To the authors' best knowledge, \cite{SchJunHac:2020:vrd} is the only contribution with a numerical relaxation in the multidimensional case.
There, a macroscopic averaged response is obtained by emulating a representative volume element under certain assumptions which yields a convex envelope per time step, thereby violating the compatibility conditions.
On the contrary, semiconvex relaxation does not violate the compatibility condition; however, the construction of a semiconvex envelope is a sophisticated task.
The construction of a semiconvex envelope can be done analytically or numerically.
The former is hard to realize for complex material laws and, thus, only few results are known as e.g.~St.-Venant--Kirchhoff elasticity \cite{LeRao:1995:qes} on $\mathbb{R}^{d\times d}$ (does not constrain $\det \boldsymbol{F} > 0)$, single slip plasticity \cite{ConThe:2005:sem,ConDolKre:2015:vms,Con:2006:rss,ConHauOrt:2007:cmc,CarConOrl:2008:man} and nematic elastomers \cite{DeSDol:2002:mrn}.
The latter can be approached in diverse ways either through enforcing laminate patterns in microstructural computations, cf.~\cite{CarPle:1997:nss,BarCarHacHop:2004:erm,KumVidKoc:2020:ant}, or by solving several optimization problems.
This contribution addresses the higher dimensional numerical rank-one relaxation in direct extension to \cite{Bar:2004:lca,Dol:1999:ncr,DolWal:2000:ena} embedded in finite element simulations to show mesh insensitivity and properties of the obtained rank-one envelope which may enable the description of strain softening.
The information tracked in the iterative convexification procedure is used to give a microstructural interpretation in terms of damage evolution.

This article is structured as follows.
Section \ref{sec:DamageModel} discusses the \((1-D)\) approach and the resulting pseudo-time-incremental damage model.
Afterwards, the notions of semiconvexity and the numerical relaxation via rank-one convexification, especially the algorithmic aspects, are discussed in Sections \ref{sec:RankOne} and \ref{sec:RankOneAlgorithm}.
Further, some general implementation aspects are described in Section \ref{sec:Implementation} and followed by the numerical simulation of boundary-value problems which show the mesh insensitivity as well as in-depth studies of the convexifications and their interpretation in terms of microstructural damage in Section \ref{sec:examples}.
We conclude in Section \ref{sec:conclusion} with some remarks on the results and future research topics, especially in the context of the efficiency regarding the numerical convexification.

%

\section{Incremental damage modeling}\label{sec:DamageModel}
We first introduce notation and basic concepts of continuum mechanics. We consider a physical body (connected, Lipschitz boundary) which in reference configuration is denoted by $\mathcal{B} \subset \mathbb{R}^3$ with coordinates $\boldsymbol{X}\in \mathcal{B}$. 
The nonlinear deformation map ${\varphi}_t : \mathcal{B} \rightarrow \mathcal{B}_t:={\varphi}_t(\mathcal{B})$ describes the deformation of the body relative to its reference configuration for times $t\in \mathbb{R}^+$.
The deformation gradient is therefore given by $\boldsymbol{F}(\boldsymbol{X}) = \text{Grad} \ {\varphi}_t(\boldsymbol{X})$ and we consider the right Cauchy--Green tensor $\boldsymbol{C}=  \boldsymbol{F}^T\boldsymbol{F}$ as deformation measure. Isotropic hyperelastic strain energy densities can be formulated in terms of the first, second and third principal invariants
\begin{equation}
	I_1 = \text{tr}\ \boldsymbol{C}, \qquad I_2 = \text{tr} \left(\text{Cof} \ \boldsymbol{C}\right), \qquad I_3 = \det \boldsymbol{C}
\end{equation}
of the Cauchy--Green tensor. The corresponding strain energy densities of virtually undamaged materials can then be phrased in the format
\begin{align*}
	\psi^0(\boldsymbol{C}) = \psi^0(I_1,I_2,I_3).
\end{align*}
As examples we utilize the St.~Venant--Kirchhoff
\begin{equation} \label{eq:st-VK}
	\psi_{\text{St.V--K}}^0(\boldsymbol{C}) \coloneqq \frac{\lambda}{8} \left(I_1 - 3\right)^2 + \frac{\mu}{4}\left(I_1^2 - 2I_1 - 2 I_2 + 3\right)
\end{equation}
and the compressible Neo-Hookean effective strain energy density
\begin{equation} \label{eq:neo-Hookean}
	\psi_{\text{NH}}^0(\boldsymbol{C}) \coloneqq \frac{\mu}{2} (I_1 - 3) - \mu \ln(J) + \frac{\lambda}{2} \ln(J)^2
\end{equation}
with the Lam\'e parameters $\mu, \lambda$ \(\in \R\) and $J = \sqrt{I_3}$.
In presence of damage, the modified strain energy density
\begin{equation}
\psi(\boldsymbol{C}, \beta) = (1-D(\beta)) \, \psi^0(\boldsymbol{C}) 
\end{equation}
is often considered, see e.g. \cite{Mie:1995:dcd}. 
Here, $\beta$ is an internal variable and evolves according to the discontinuous damage approach 
\begin{equation}
	\beta_t \coloneqq \max_{s\leq t} \left[ \psi^0(\boldsymbol{C}_s)\right] \quad \text{for } s,t \in \mathbb{R}^+
\end{equation}
and $D: \R \to [0,1)$ is the non-decreasing damage function of the form 
\begin{equation}
	D(\beta) = D_{\infty} \left(1-\exp\left(- \frac{\beta}{D_0}\right)\right),
\end{equation}
where $D_{\infty}\in (0,1)$ is the asymptotic limit of the damage function and $D_0\in \R^+$ is the damage saturation parameter.

To relax the phenomenological scalar-valued continuum damage mechanics energy density, we briefly recapitulate the derivation of the associated incremental stress potential as obtained in \cite{BalOrt:2012:riv}.
Let \(t_0, \ldots, t_N\) denote a time discretization of the time interval \([0, T]\) and \(\Delta t_{k+1} := t_{k+1} - t_k\) denote the incremental time steps.
Furthermore, let $\dot{\psi}$ denote the time derivative of strain energy density and $\phi$ the dissipation potential.
According to \cite{OrtRep:1999:nem}, the generalized work done in the body within the time increment \(\Delta t_{k+1}\) is then given by
\begin{equation}
	\mathcal{W}(\boldsymbol{F}_{k+1},\beta_{k+1}) \coloneqq \int_{t_k}^{t_{k+1}} \dot{\psi} + \phi \ \text{d}t.
\end{equation}
The incremental stress potential
\begin{equation}
	\label{eq:IVF}
	W(\boldsymbol{F}_{k+1}) = \inf_{\beta_{k+1}} \left[\mathcal{W}(\boldsymbol{F}_{k+1},\beta_{k+1})\right],
\end{equation}
which only depends on the deformation gradient $\boldsymbol{F}_{k+1}$, minimizes $\mathcal{W}$ with respect to the internal variable $\beta_{k+1}$ in each incremental time step.
Via the second law of thermodynamics, one can derive that the dissipation potential is of the form $\phi \coloneqq \psi^0\dot{D}$. 
With this in mind, the analytical integration and minimization of \eqref{eq:IVF} yields 
\begin{equation}\label{eq:ISP}
	W(\boldsymbol{F}) = \psi(\boldsymbol{F},D) - \psi(\boldsymbol{F}_k,D_k)+\beta D - \beta_k D_k - \overline{D} + \overline{D}_k.
\end{equation}
For a more detailed derivation, the reader is referred to \cite{BalOrt:2012:riv}.
Here, \(\overline{D}\) denotes the antiderivative of the damage function.
For an improved readability, the time-step dependence was shortened by dropping the index $k+1$ and the abbreviations $D\coloneqq D(\beta)$, $D_k \coloneqq D(\beta_k)$, $\overline{D} \coloneqq \overline{D}(\beta)$, and $\overline{D}_k \coloneqq \overline{D}_k(\beta)$ denote evaluations of the function $D$ and its antiderivative $\overline{D}$ at the minimizers $\beta$ of the current time step and the values of the internal variable at the last time step $\beta_k$.

Let the potential energy of external forces $\hat{{t}}$, which are applied at the Neumann boundary \(\partial \mathcal{B}_\sigma\), be denoted by $\Pi^{\text{ext}}$.
Further, let \(\partial B_\varphi\) denote the Dirichlet boundary and \(\hat{{\varphi}}\) the Dirichlet boundary conditions on \(\partial B_\varphi\). 
We consider the total potential energy
\begin{equation*}
	\Pi({\varphi}) \coloneqq \int_{\mathcal{B}} W(\boldsymbol{F}({\varphi})) \ \text{d}V + \Pi^{\text{ext}}(\hat{{t}},{\varphi}).
\end{equation*}

The goal is now to follow the principle of minimum potential energy to obtain the state of mechanical equilibrium and to find minimizers of this energy subject to the boundary conditions, i.e.
\begin{equation}
\inf_{{\varphi}} \left\{\Pi({\varphi})\;|\; {\varphi}=\hat{{\varphi}} \text{ on } \partial\mathcal{B}_{\varphi}, {t}=\hat{{t}} \text{ on } \partial\mathcal{B}_{\sigma}\right\}.
\end{equation}
Setting the first variation of $\Pi$ to zero leads to the weak form in the total Lagrangian setting
\begin{equation}
\label{eq:weakform}
\int_{\mathcal{B}} \boldsymbol{P} : \text{Grad} \delta \boldsymbol{u} \text{ d}V - \int_{\partial\mathcal{B}_{\sigma}} \hat{t} \cdot \delta \boldsymbol{u} \text{ d}S = 0,
\end{equation}
subject to Dirichlet boundary conditions $\boldsymbol{u} = \hat{\boldsymbol{u}}$ on $\partial \mathcal{B}_{{\varphi}}$, see e.g.~\cite[Sec. 8.3]{Hol:2000:nsm} for a detailed derivation.
The first Piola--Kirchhoff stress tensor $\boldsymbol{P}$ and its derivative, the nominal tangent moduli $\mathbb{A}$, are computed as
\begin{equation}
\boldsymbol{P} = \frac{\partial W(\boldsymbol{F})}{\partial \boldsymbol{F}}, \qquad \mathbb{A} = \frac{\partial \boldsymbol{P}}{\partial \boldsymbol{F}} = \frac{\partial^2 W(\boldsymbol{F})}{\partial \boldsymbol{F} \partial \boldsymbol{F}}.
\end{equation}

The problem \eqref{eq:weakform} can be solved numerically by using the finite element method with a suitable solver for the resulting system of nonlinear equations.
For a steepest descent scheme like \cite[Algorithm 9.1]{Bar:2015:nmn}, only the first Piola--Kirchhoff stresses $\boldsymbol{P}$ are required, whereas for second order methods, such as the Newton method (cf.~\cite[Algorithm 4.2]{Bar:2015:nmn}), the derivative of these, the tangent moduli $\mathbb{A}$, need to be computed or approximated additionally. 

Due to the dissipative nature of damage evolution processes, the time-incremental energy density \eqref{eq:ISP} becomes non-convex. Therefore, the direct numerical simulation of this problem suffers from stability issues and mesh dependency.

%

\section{Relaxation by rank-one convexification} \label{sec:RankOne}
To treat the drawbacks resulting from non-convexity, we apply a relaxation approach and replace the energy density $W$ by a (semi)convex envelope. Possible choices are the convex, polyconvex, quasiconvex and rank-one convex envelopes. Their relations are described by the inequality chain 
$$W^{\text{c}} \leq W^{\text{pc}} \leq W^{\text{qc}} \leq W^{\text{rc}} \leq W.$$

Replacing $W$ by the convex envelope $W^{\text{c}}$ may lead to nonphysical material behavior and is therefore of limited relevance in continuum mechanics models, cf. \cite[Theorem 4.8-1]{Cia:1988:mea}.
The polyconvex hull is less restrictive but presumably an overly high computational effort can be expected for computational convexification.
For the quasiconvex envelope, which would be the hull of choice from a theoretical perspective, no efficient algorithm is known.
Therefore, our choice is the rank-one convex envelope since it delivers a reasonable approximation to the quasi-convex envelope and provides a physical interpretation in terms of laminated microstructures.
Moreover, we will show that this choice is able to capture strain softening to some extent.

\subsection{Definition and alternative characterizations}
A function \(W: \R^{d \times d} \to \R\) is called \emph{rank-one convex}, if 
\begin{equation}
	W(\lambda \, \boldsymbol{A} + (1 - \lambda) \, \boldsymbol{B}) \leq \lambda \, W(\boldsymbol{A}) + (1 - \lambda) \, W(\boldsymbol{B})
\end{equation}
holds for all \(\lambda \in [0, 1]\) and all matrices \(\boldsymbol{A}, \boldsymbol{B} \in \R^{d \times d}\) that are rank-one connected, i.e.~\(\rank(\boldsymbol{A} - \boldsymbol{B}) = 1\).  This condition is equivalent to the property that \(W\) is convex along rank-one directions, i.e.~the function \(g:\eta \mapsto W(\boldsymbol{F} + \eta \, \boldsymbol{a} \otimes \boldsymbol{b})\) is convex for all \(\boldsymbol{a}, \boldsymbol{b} \in \R^{d}\) and \(\boldsymbol{F} \in \R^{d \times d}\). 
The \emph{rank-one convex envelope} \(W^{\text{rc}}\) of a function \(W\) is the largest rank-one convex function below \(W\), i.e. 
\begin{equation}
	W^{\text{rc}}(\boldsymbol{F}) = \sup\{\tilde{W}(\boldsymbol{F}) \mid \tilde{W}: \R^{d \times d} \to \R \text{ with } \tilde{W} \leq W,  \,\tilde{W} \text{ is rank-one convex}\}.
\end{equation}

Various characterizations for the rank-one convex hull can be found for example in \cite[Chapter 5]{Dac:2008:dmc}.
The first example is the construction by successive lamination.
This is done by setting \(W^{0} = W\) and for \(k > 0\) the next laminate is obtained by
\begin{align}
	W^{k + 1} (\boldsymbol{F}) & = \inf_{\substack{\lambda \in [0, 1] \\ \boldsymbol{A}, \boldsymbol{B} \in \mathbb{R}^{d \times d}}}\left\{ \lambda \, W^{k}(\boldsymbol{A}) + (1 - \lambda) \, W^{k}(\boldsymbol{B}) \,  \bigg\vert 
	\begin{array}{c} 
	\lambda \, \boldsymbol{A} + (1 - \lambda) \, \boldsymbol{B} = \boldsymbol{F}, \\ \mathrm{rank}(\boldsymbol{A} - \boldsymbol{B}) = 1
	\end{array} \right\}
\end{align}
at all points \(\boldsymbol{F} \in \R^{d \times d}\). The lamination-convex envelope \(W^{\text{lc}}\) is defined in a pointwise manner through the limit 
\begin{equation*} 
	W^{\text{lc}} (\boldsymbol{F}) \coloneqq \lim\limits_{k \to \infty} W^{k} (\boldsymbol{F}).
\end{equation*}
In \cite[5.C]{KohStr:1986:odr}, it is shown that the rank-one convex envelope can be obtained by successive lamination, that is \(W^{\text{rc}} = W^{\text{lc}}\).
This characterization of the rank-one convex envelope is exploited in the algorithmic computations below.

An alternative characterization links the rank-one characterization via the successive lamination to so called \(\mathcal{H}_M\) sequences.  
We utilize the notation from \cite{Dol:1999:ncr}.
Shorthand, \(\mathcal{H}_M\) is a condition on a set of matrices \(\boldsymbol{F}_{1}, \dots,\boldsymbol{F}_{M}\) that they are in some sense hierarchically rank-one connected, the mathematically rigorous definition is given in the following.
The sequence of tuples \((\xi_i, \boldsymbol{F}_i) \in [0, 1] \times \R^{d \times d}\) for \(i = 1, \dots, M\) is said to be an \(\mathcal{H}_M\) sequence (notation: \((\xi_i, \boldsymbol{F}_i) \in \mathcal{H}_M\)) if \(\sum_{i = 1}^{M} \xi_i = 1\) and the following condition holds:
\begin{itemize}[leftmargin=0.7cm]
	\item in case \(M = 2\) it holds \(\rank(\boldsymbol{F}_1 - \boldsymbol{F}_2) = 1\);
	\item in case \(M > 2\) up to a permutation of the indices \(\{1, \dots, M\}\) it holds {\({\rank(\boldsymbol{F}_1 - \boldsymbol{F}_2) = 1}\)} and for all \(i \in \{2, \dots, M-1\}\) define
	\begin{align*}
		\mu_1 & = \xi_1 + \xi_2, &\hat{\boldsymbol{F}}_{1} & = \frac{1}{\mu_1} (\xi_1 \boldsymbol{F}_1 + \xi_2 \boldsymbol{F}_2), \\
		\mu_i & = \xi_{i+1}, & \hat{\boldsymbol{F}}_i & = \boldsymbol{F}_{i+1}
	\end{align*}
	and it holds \((\mu_i, \hat{\boldsymbol{F}}_i) \in \mathcal{H}_{M - 1}\). 
\end{itemize}
With this definition at hand the rank-one convex envelope can be characterized by
\begin{equation}
	W^{\text{rc}}(\boldsymbol{F}) = \inf_{} \left\{\sum_{i = 1}^{M} \xi_i \, W(\boldsymbol{F}_i) \, \Big\vert \, M \in \N, (\xi_i, \boldsymbol{F}_i) \in \mathcal{H}_M , \boldsymbol{F} = \sum_{i = 1}^{M} \xi_i \boldsymbol{F}_i\right\}
\end{equation}
where \(M \in \N\) cannot be bounded in general. The equivalence of the characterization is shown in \cite[Proposition 5.16]{Dac:2008:dmc}.

%

\section{Numerical rank-one convexification} \label{sec:RankOneAlgorithm}
This section presents algorithms for the numerical approximation of rank-one convex envelopes along with micro-mechanical interpretations in terms of iterative rank-one lamination.

\subsection{The basic algorithm}
We begin by recalling the important aspects of the algorithm presented in \cite{Bar:2004:lca} and \cite[Chapter 9.3.4]{Bar:2015:nmn}. By \(|\cdot|_\infty\) we denote the component-wise maximum norm of a vector or a matrix, e.g. \(|\boldsymbol{A}|_\infty = \max_{ij} |\boldsymbol{A}_{ij}|\).
We assume a discretization of the energy density \(W:\R^{d \times d} \to \R\) as a continuous piecewise linear approximation \(W_{\delta, r}\) on a mesh
\begin{equation} \label{eq:mesh}
	\mathcal{N}_{\delta, r} 
	= \delta \, \Z^{d \times d} \cap \left\{\boldsymbol{A}\in \R^{d \times d} \,\big\vert\, |\boldsymbol{A}|_\infty \leq r\right\}
\end{equation}
with parameters for the mesh width \(\delta \in \R\) and the radius of the ball \(r \in \R\). The radius needs to be chosen sufficiently large so that the mesh is able to capture all relevant minima contributing to the convexification. Further, a discrete set of rank-one directions is considered 
\begin{equation} \label{eq:bartelsdirections}
	\mathcal{R}^1_{\delta, r} = \{\boldsymbol{a} \otimes \boldsymbol{b} \mid \boldsymbol{a}, \boldsymbol{b} \in \delta \, \Z^d , \|\boldsymbol{a}\|_\infty \leq 2\, d\, r, 1 - d \, \delta \leq \|\boldsymbol{b}\|_\infty \leq 1 + d \, \delta\}.
\end{equation}
For the sake of readability, we denote the discretized rank-one set by \(\mathcal{R}\).
Set \(W_{\delta, r}^{0}(\boldsymbol{F}) = W(\boldsymbol{F})\) for all \(\boldsymbol{F} \in \mathcal{N}_{\delta, r}\). The iterative lamination is now performed in a pointwise manner. Therefore, for all \(\boldsymbol{F} \in \mathcal{N}_{\delta, r}\) the optimization problem 
\begin{equation} \label{eq:minglobiter}
	W_{\delta, r}^{k+1} (\boldsymbol{F}) = \inf_{} 
	\left\{\lambda \, W^{k}_{\delta, r} (\boldsymbol{F} + \delta l_1 \boldsymbol{R}) + (1 - \lambda) \, W^{k}_{\delta, r} (\boldsymbol{F} + \delta l_2 \boldsymbol{R}) \, \Big\vert \, 
	\substack{\boldsymbol{R} \in \mathcal{R}, \, \lambda \in [0,1], \, l_1, l_2 \in \Z \\
	\lambda \, l_1 + (1 - \lambda) \, l_2 = 0}
	\right\}
\end{equation}
is solved. Function values corresponding to points that are not represented in the mesh \(\mathcal{N}_{\delta, r}\) are obtained by linear interpolation.
The constraint on \(l_1, l_2\) and \(\lambda\) in the above optimization problem ensures that the convex combination of the arguments of \(W_{\delta, r}^{k}\) leads to \(\boldsymbol{F}\).
The iteration is stopped until either a number of maximal iterations \(k_{\text{max}}\) is reached or the change in two consecutive iterates is small enough, i.e. \(|W^{k + 1}_{\delta, r} - W^{k}_{\delta, r}|_\infty \leq \text{tol}\).

The solution of the minimization problem can be performed by considering for every rank-one direction \(\boldsymbol{R}\in\mathcal{R}\) the one-dimensional convexification along the line \(\boldsymbol{F} + \eta \, \boldsymbol{R}\) for \(\eta \in \R\). 
We adapt the notation of \cite{DolWal:2000:ena} and denote by \(\ell_{\delta}(\boldsymbol{F}, \boldsymbol{R})\) the points along the rank-one direction \(\boldsymbol{R} \in \mathcal{R}\) through the point \(\boldsymbol{F}\), i.e.
\begin{align} \label{eq:rankOneLine}
	\ell_{\delta}(\boldsymbol{F}, \boldsymbol{R}) = \left\{\boldsymbol{F} + l \,\delta \,\boldsymbol{R} \mid l \in \Z \right\} \cap \mathrm{conv}(\mathcal{N}_{\delta, r}).
\end{align}

For fixed iteration \(k\) and fixed mesh point \(\boldsymbol{F}\in \mathcal{N}_{\delta, r}\) as well as direction \(\boldsymbol{R} \in \mathcal{R}\), let \(w_l = W_{\delta, r}^{k}(\boldsymbol{F} + l\,\delta \,\boldsymbol{R})\) denote the function values of \(W^{k}_{\delta, r} (\ell_{\delta}(\boldsymbol{F}, \boldsymbol{R}))\) and \(I_\ell\) the index set of all \(l\) of \eqref{eq:rankOneLine}.
Since the matrices in \(\ell_\delta(\boldsymbol{F}, \boldsymbol{R})\) are not directly represented in the mesh in general, the function values of these intermediate points are obtained by interpolation.
The mapping \(q: I_\ell \to W^{k}_{\delta, r}(\ell_{\delta}(\boldsymbol{F}, \boldsymbol{R}))\) can now be interpreted as a one-dimensional function, mapping each index \(l \in I_\ell\) to the value \(w_l\).
This one-dimensional function needs now to be convexified. 
This allows for the application of the convexification procedure presented in Algorithm \ref{alg:oneDimConvexification}. 
Afterwards the minimum over all \(\boldsymbol{R} \in \mathcal{R}\) is stored as the resulting relaxed value for the mesh point \(\boldsymbol{F}\) in the current iteration.

\begin{figure}
\begin{subfigure}[b]{0.48\textwidth}
	\centering
	\scalebox{0.48}{\input{figures/convexify2d.pgf}}
	\caption{Mesh \(\mathcal{N}_{\delta, r}\) (blue) and discretized direction set (orange) in a single mesh point}
	\label{fig:RankOneConvexificationMesh}
\end{subfigure}
\hfill
\begin{subfigure}[b]{0.48\textwidth}
	\centering
	\scalebox{0.48}{\input{figures/convexify.pgf}}
	\caption{Convexification of the one dimensional mapping {\({I_\ell \to W^{k}_{\delta, r}(\ell_{\delta}(\boldsymbol{F}, \boldsymbol{R}))}\)} (gray) by Algorithm \ref{alg:oneDimConvexification} delivers the green hull} 
	\label{fig:RankOneConvexification1D}
\end{subfigure}
\caption{Rank-one convexification. \subref{fig:RankOneConvexificationMesh}: schematic representation of the deformation gradient mesh \(\mathcal{N}\) (projected to two dimensions) and  discretized rank-one directions (orange lines) through point \(F\).
\subref{fig:RankOneConvexification1D}: one-dimensional convexification of the function based on the interpolations from \subref{fig:RankOneConvexificationMesh}. 
Original function is drawn in light blue, piece-wise linear interpolation in gray, resulting one-dimensional convex hull in green, the function value for \(F\) (red) is obtained by linear interpolation, which is given by the convex combination of function values corresponding to \(F^+\) and \(F^-\).}
\label{fig:RankOneConvexification}
\end{figure}

\subsection{One-dimensional convexification} \label{sec:one-dim-convexification}
Let \(L\) be the number of matrices in the set \(\ell_{\delta}(\boldsymbol{F}, \boldsymbol{R})\), i.e. \(L \coloneqq |\ell_{\delta}(\boldsymbol{F}, \boldsymbol{R})|\), and assume that the distribution of points along the rank-one line is equidistant with distance \(\delta\). 
Let the \(x_i \in \Z\) for \(i = 0, \ldots, L\) denote the indices in the set \(I_\ell\).
The convexification of the mapping \(x_i \mapsto w_i\) (a shifted/renumbered version of \(q\)) can then be realized through the following procedure described in Algorithm \ref{alg:oneDimConvexification}.

\centerline{
\begin{minipage}{0.93\textwidth}
\begin{algorithm}[H]
	\begin{algorithmic}[1]
		\State{\textbf{Input:} \texttt{x, w} \Comment{arrays of length $L$}}
		\State{\textbf{Output:} \texttt{y, c} \Comment{arrays of length $n$}}
		\State{\texttt{y[1] = w[1], y[2] = w[2], c[1] = w[1], c[2] = w[2]}}
		\State{\texttt{n = 2}}
		\For{\texttt{i = 3, 4, \ldots, L}}
		\While{\texttt{(c[n] - c[n-1]) * (x[i] - y[n]) >=}\\
		\texttt{\qquad\qquad\qquad\qquad(w[i] - c[n]) * (y[n] - y[n-1])} \textbf{and} \texttt{n >=  1}}
		\State{\texttt{n -= 1}}					
		\EndWhile{}
		\State{\texttt{n += 1}}
		\State{\texttt{y[n] = x[i], c[n] = w[i]}}
		\EndFor{}
		\State \textbf{return} \texttt{y, c}
	\end{algorithmic}
	\caption{\texttt{convexify(x, w)}}
	\label{alg:oneDimConvexification}
\end{algorithm}
\end{minipage}
}~\\

This algorithm realizes the convexification by iteration from left to right of the interval, successively computing difference quotients and checking two adjacent slopes for convexity. 
Depending on the angle of the two slopes, either the points already visited are deleted (by overwriting \texttt{y[n]} and \texttt{c[n]}) or the new point is added (line 9).
The algorithm realizes the convexification in linear complexity in the number of input points \(L\), due to the deletion of points already visited and not contributing to the convex hull.
The resulting arrays of length \(n\) contain the supporting points of the convex hull.
In computational geometry, this procedure is also known as Graham's scan \cite{Gra:1972:ead}.
An adaptive convexification strategy, which accelerates the convexification significantly by using second order derivative information, can be found in \cite{KohNeuMelPetPetBal:2022:acm}.
After the convexification the function value of the convex hull corresponding to \(\boldsymbol{F}\) is obtained by interpolation of the function \(y \to c\) at the index \(x_i \in {0, \dots, L}\) corresponding to the point \(l = 0\).

The solution of the optimization problem \eqref{eq:minglobiter} and hence one global iteration \(k \to k +1\) in the rank-one convexification algorithm is of complexity \(\mathcal{O}(N^{d^2 + 2 \, d + 1})\), where \(N\) denotes the number of discretization points in one matrix component in the mesh.
The overall number of performed iterations can be interpreted as the maximum lamination depth of the microstructure. 

\begin{remark}[An alternative algorithm by \cite{Dol:1999:ncr, DolWal:2000:ena}]
While the above algorithm originally introduced by \cite{Bar:2004:lca, Bar:2015:nmn} only alters the function value of a single mesh point in each one-dimensional convexification, the algorithm presented in \cite{Dol:1999:ncr, DolWal:2000:ena} considers all mesh points along a specified rank-one line and updates all of them after the one-dimensional convexification; the characterization of the discrete set of rank-one direction matrices is therefore different. 
The sets of rank-one directions considered are characterized by
\begin{align} \label{eq:dolzmanDirections}
	\mathcal{R}^1_{k} = \{\delta \, \boldsymbol{a} \otimes \boldsymbol{b} \mid \boldsymbol{a},\boldsymbol{b} \in \Z^d, |\boldsymbol{a}|_\infty,|\boldsymbol{b}|_\infty \leq k\}
\end{align}
with \(k \in \N\).
Due to the update of function values for all mesh points along the rank-one line, the one-dimensional convexifications can not be carried out over all points simultaneously.
However, for a fixed rank one direction parallelization is possible by running lines in parallel through the mesh.
\end{remark}

\subsection{Rank-one lamination tree} 
\begin{figure}
	\centering
	\ifthenelse{\boolean{professormode}}
	{\includegraphics{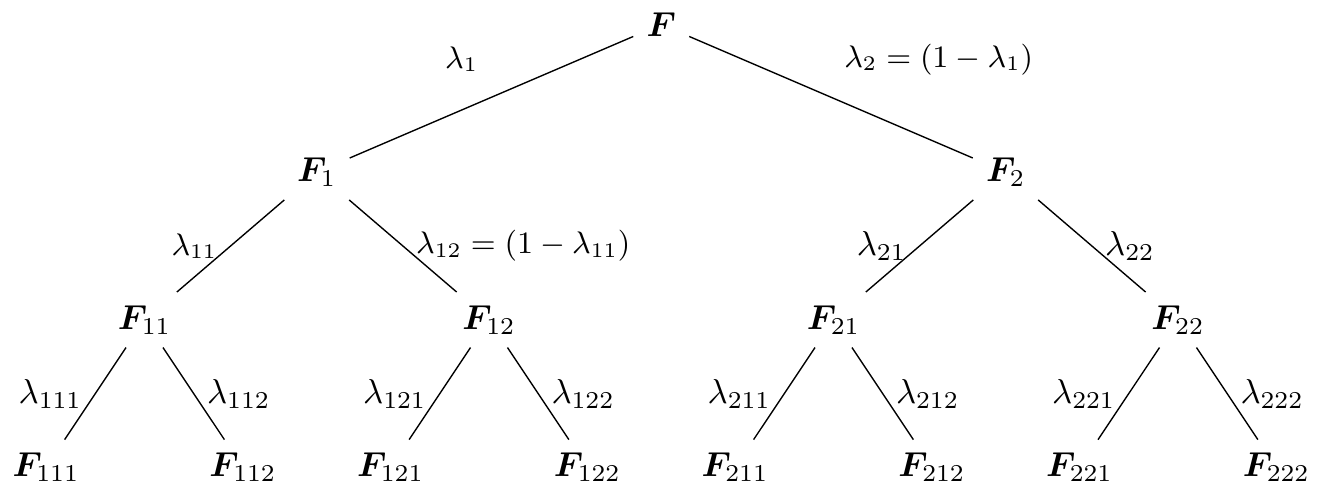}}
	{\tikzsetfigurename{laminationtree}
\begin{tikzpicture}
	\pgfmathsetmacro\firstsize{3.5}
	\pgfmathsetmacro\secondsize{1.75}
	\pgfmathsetmacro\thirdsize{1.0}
	\pgfmathsetmacro\verticalsize{1.5}
	\node (F) at (0,0) {\small 
		$\boldsymbol{F}$};
	\node (F1) at (-\firstsize,-\verticalsize) {\small
		$\boldsymbol{F}_1$};
	\node (F2) at (\firstsize,-\verticalsize) {\small
		$\boldsymbol{F}_2$};
	\draw (F) -- (F1) node[midway,left,yshift=1em] {\footnotesize $\lambda_1$};
	\draw (F) -- (F2) node[midway,right,yshift=1em] {\footnotesize $\lambda_2 = (1 - \lambda_1)$};
	\node (F11) at (-\firstsize-\secondsize,-2*\verticalsize) {\small
		$\boldsymbol{F}_{11}$};
	\node (F12) at (-\firstsize+\secondsize,-2*\verticalsize) {\small
		$\boldsymbol{F}_{12}$};
	\draw (F1) -- (F11) node[midway,left] {\footnotesize $\lambda_{11}$};
	\draw (F1) -- (F12) node[midway,right] {\footnotesize $\lambda_{12} = (1-\lambda_{11})$};
	\node (F21) at (\firstsize-\secondsize,-2*\verticalsize) {\small
		$\boldsymbol{F}_{21}$};
	\node (F22) at (\firstsize+\secondsize,-2*\verticalsize) {\small
		$\boldsymbol{F}_{22}$};
	\draw (F2) -- (F21) node[midway,left] {\small $\lambda_{21}$};
	\draw (F2) -- (F22) node[midway,right] {\small $\lambda_{22}$};
	\node (F111) at (-\firstsize-\secondsize-\thirdsize,-3*\verticalsize) {\small
		$\boldsymbol{F}_{111}$};
	\node (F112) at (-\firstsize-\secondsize+\thirdsize,-3*\verticalsize) {\small
		$\boldsymbol{F}_{112}$};
	\draw (F11) -- (F111) node[midway,left] {\small $\lambda_{111}$};
	\draw (F11) -- (F112) node[midway,right] {\small $\lambda_{112}$};
	\node (F121) at (-\firstsize+\secondsize-\thirdsize,-3*\verticalsize) {\small
		$\boldsymbol{F}_{121}$};
	\node (F122) at (-\firstsize+\secondsize+\thirdsize,-3*\verticalsize) {\small
		$\boldsymbol{F}_{122}$};
	\draw (F12) -- (F121) node[midway,left] {\small $\lambda_{121}$};
	\draw (F12) -- (F122) node[midway,right] {\small $\lambda_{122}$};
	\node (F211) at (+\firstsize-\secondsize-\thirdsize,-3*\verticalsize) {\small
		$\boldsymbol{F}_{211}$};
	\node (F212) at (+\firstsize-\secondsize+\thirdsize,-3*\verticalsize) {\small
		$\boldsymbol{F}_{212}$};
	\draw (F21) -- (F211) node[midway,left] {\small $\lambda_{211}$};
	\draw (F21) -- (F212) node[midway,right] {\small $\lambda_{212}$};
	\node (F221) at (+\firstsize+\secondsize-\thirdsize,-3*\verticalsize) {\small
		$\boldsymbol{F}_{221}$};
	\node (F222) at (+\firstsize+\secondsize+\thirdsize,-3*\verticalsize) {\small
		$\boldsymbol{F}_{222}$};
	\draw (F22) -- (F221) node[midway,left] {\small $\lambda_{221}$};
	\draw (F22) -- (F222) node[midway,right] {\small $\lambda_{222}$};
	
\end{tikzpicture}}
	\caption{A rank-one tree obtained by successive lamination (for iterate \(k = 3\)) of a function $W$ at the point \(\boldsymbol{F}\). Note that only branching due to lamination and not due to interpolation is illustrated. 
	The \(\boldsymbol{F}_1\) and \(\boldsymbol{F}_2\) can be identified with \(\boldsymbol{F}^-\) and \(\boldsymbol{F}^+\). 
	In the deeper parts of the tree \(\boldsymbol{F}_{121}\) and \(\boldsymbol{F}_{122}\) are for example associated to \(\boldsymbol{F}^{-}_{12}\) and \(\boldsymbol{F}^{+}_{12}\), due to the branching of \(\boldsymbol{F}_{12}\).
	In general, the rank-one tree can be unbalanced due to bifurcation only in one subtree. 
	}
	\label{fig:laminationtree}
\end{figure}
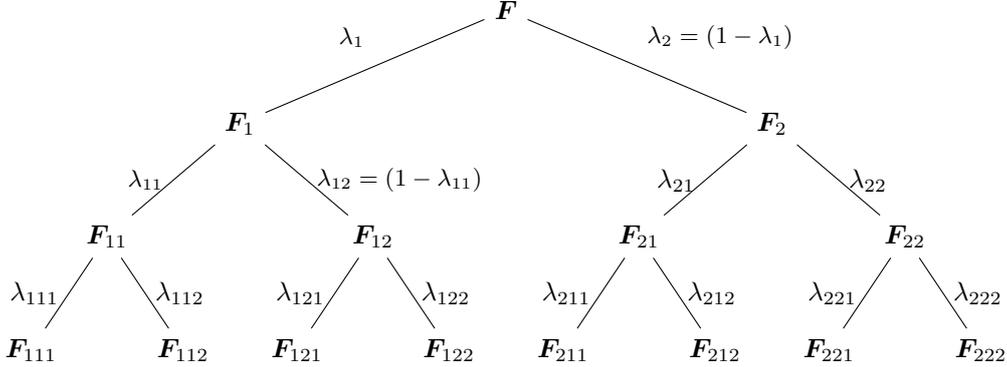

The consecutive optimization problems \eqref{eq:minglobiter} only require the function values of the iterates \(W^{k}_{\delta, r}\) for all mesh points.
Deeper insights into the structure of the hull can be gained if further knowledge is tracked. 
This is possible by considering the hierarchical construction of the \(\mathcal{H}_M\) sequences.
By tracking the solution arguments of the optimization problem \eqref{eq:minglobiter}, it is possible to set up a lamination tree for each point \(\boldsymbol{F}\) that is convexified, as illustrated in Figure \ref{fig:laminationtree}.
For each convexified node \(\boldsymbol{F}\), a child is appended for \(\boldsymbol{F}^+\) and \(\boldsymbol{F}^-\) and their associated convex coefficient \(\lambda\) is tracked as weight over the edges. 
After each global iteration \(k\), the tree has to be updated in terms of the new minimizers and their subtrees (see Algorithm \ref{alg:tree} for details).
A slight adjustment is required if the tree is computed numerically; the points (gray points on the orange rank-one  line in the left picture of Figure \ref{fig:RankOneConvexification}) \(\boldsymbol{F}^+ = \boldsymbol{F} + \delta l_1 \boldsymbol{R}\) and \(\boldsymbol{F}^- = \boldsymbol{F} + \delta l_1 \boldsymbol{R}\) (where \(\boldsymbol{R}, l_1, l_2\) denote the minimizing arguments of \eqref{eq:minglobiter}) are in general not represented in the mesh. 
Due to this aspect, one has to consider the nearest neighbours of these points that contribute to the linear interpolation.
Hence the rank-one tree contains branching subject to the interpolation (16 children) and subject to the rank-one lamination (two children). 

Even if the setup of the tree seems straightforward from an iterative perspective, the tree can change substantially in one iteration, due to the occurrence of different minima corresponding to new lamination distributions. 
Hence, whole subtrees are subject to change in an iteration.

The tree can be used to approximate derivatives (or in general subdifferentials) of the convex hull by linear combination of the leaves of the tree.
Those leaves correspond to the last branching, i.e.~the case \(\mathcal{H}_2\).
Examples of lamination trees and the microstructural interpretation will follow in Section \ref{sec:examples}.

%

\section{Complexity reduction and efficient implementation}\label{sec:Implementation}
While the convexification algorithms appear promising on paper, the main challenge is the overall computational workload as outlined in Section \ref{sec:one-dim-convexification} above. 
To make the rank-one convexification computationally feasible, this section presents various adjustments that significantly accelerate the previous algorithms. In addition, the nontrivial extraction of derivative information from the computed convex hull for the optimization algorithms is addressed. 

\subsection{Reduction of rank-one lines}
Originally, the discretization of the rank-one directions, the set \(\mathcal{R}_{\delta, r}^1\), is coupled to the discretization of the deformation gradient space, the mesh \(\mathcal{N}_{\delta, r}\). 
This coupling is motivated by the theoretical convergence proof presented in \cite{Bar:2004:lca, Bar:2005:rea}. 
However, the huge set of directions may be pessimistic from a practical perspective.
Our numerical experiments show that a reduced set of directions suffices to achieve computationally feasible simulations of simplistic boundary value problems.
Therefore, we use the discrete direction set \(\mathcal{R}^1_{k}\) as defined in \eqref{eq:dolzmanDirections} for \(k = 1\) as proposed in \cite[Sec. 5.1]{DolWal:2000:ena}. 
This results in the set 
\begin{align} \label{eq:redrankonedir}
	\mathcal{R}^1_{1} & = \{\delta \, \boldsymbol{a}\otimes \boldsymbol{b} \mid \boldsymbol{a}, \boldsymbol{b} \in \Z^d, |\boldsymbol{a}|_\infty, |\boldsymbol{b}|_\infty \leq 1\}
\end{align}
consisting of $16$ distinct rank-one directions if symmetries are exploited, that is the sign of the directions is omitted.

\subsection{Adaptive discretization of the deformation-gradient space}\label{sec:adaptive}
The mesh \(\mathcal{N}_{\delta, r}\) was constructed in \eqref{eq:mesh} as an equidistant discretization, to be more precise, a hypercube with edge length \(2r\) and resolution \(\delta\) of the space \(\R^{d \times d}\). 
The mesh size parameter \(\delta\) and the radius of the hypercube were applied to all of the components of the matrix space. Letting \(N = \frac{2r}{\delta}\) denote the number of grid points in one matrix component, i.e. unit direction of the space \(\R^{d \times d}\), the mesh consists of \(N^{d^2}\) points.
To reduce the number of mesh points, we now treat each component individually. 
Instead of setting up a hypercube of edge length \(2r\) we consider points in the cuboid that is characterized by \(F_{ij}^{\text{min}} \leq F_{ij}^{\text{max}}\) for \(i, j = 1,\dots, d\). 
Furthermore, the mesh-width parameter is adjusted for each dimension, i.e. \(\delta_{ij} \in \R\).
Overall, we utilize the discrete set of points
\begin{equation*}
	\mathcal{N}_{\delta, r} = \left\{\boldsymbol{F} \in \R^{d \times d} \,\mid\, F_{ij}\in \delta_{ij}\,\Z, F_{ij}^{\text{min}} \leq F_{ij} \leq F_{ij}^{\text{max}} \right\}
\end{equation*}
as our deformation gradient grid.
For simplicity, we use the shorthand notation \(\mathcal{N}\) instead of \(\mathcal{N}_{\delta, r}\) for the deformation gradient mesh as long as there is no confusion about parameters.
To be more precise, depending on the boundary value problems and the expected occurrence of deformations, the bounds of each entry in the discretization are set accordingly and the shear components are not fully resolved in the mesh \(\mathcal{N}\) but only represented by a coarse discretization. 
This approach turns out reasonable in our numerical experiments.

\subsection{Reliable computation of derivative information}
To perform a descent method or a (quasi-)Newton scheme, first-order and second-order derivative information of the incremental stress potential is required to set up the first and second variation of the energy functional.
However, the constructed piecewise linear rank-one convex envelope is only $C^0$ and its derivative is not well defined, e.g. in the mesh points. 
With the regularity results of \cite{BalKirKri:2000:rqe} in mind, our consistent computation of derivatives utilizes the two-folded decomposition of the approximated rank-one convex hull $W^{k}_{\delta,r}(\boldsymbol{F})$ in its $\mathcal{H}_k$ sequence and, if necessary, the weighted sum of the linear interpolation to formulate a derivative in dependence on the smoother function $W(\boldsymbol{F})$, i.e.
\begin{equation}\label{eq:recursivederivative}
    \partial_F W^k_{\delta,r} (\boldsymbol{F}) = \begin{cases}
        \xi \, \partial_F W^{k-1}_{\delta,r}(\boldsymbol{F}^+) + (1-\xi) \, \partial_F W^{k-1}_{\delta,r}(\boldsymbol{F}^-) & \boldsymbol{F} \in \mathcal{N}\\
        \sum_i^{2^{d\times d}}w_i \, \partial_F W^{k}(\boldsymbol{F}_i) & \boldsymbol{F} \notin \mathcal{N}.
    \end{cases}
\end{equation}
Note that the index \(k\) only refers to the lamination/global iteration. The branching in the tree due to interpolation is only encoded in the sum of current iteration (\(k\)) laminates.
In the case $\boldsymbol{F} \in \mathcal{N}$, the matrices $\boldsymbol{F}^+$ and $\boldsymbol{F}^-$ correspond to the minimizing arguments of the optimization problem \eqref{eq:minglobiter}.
Since the linear interpolation (as handled in the case $\boldsymbol{F} \notin \mathcal{N}$) can have in the two-dimensional case a maximum of $2^{2\times 2} = 16$ nodes per cell, the tree can branch into a maximum of sixteen different children where the volume fractions correspond to the interpolation weights, i.e. the \(w_i \in [0, 1]\) for the mesh points  \(\boldsymbol{F}_i\).
Further, if a given deformation gradient $\boldsymbol{F}$ is in the set $\mathcal{N}$, the branching is bounded by two children.
However, the resulting tree is unbalanced and not structured at all, which renders a sophisticated construction task.
Thus, we present a stack (queue) based algorithm to construct the two-folded decomposition tree.
Algorithm~\ref{alg:tree} presents the procedure and works as follows:
first, the root node is constructed and the queue is initialized.
The queue consists of a tuple containing the node and its parent.
The first candidates are queued from line 6 to 14 and if the given deformation gradient $\boldsymbol{F}$ is not part of the convexification grid, it is decomposed into the weights and points that contribute to the interpolation sum.
Afterwards a two staged while loop begins.
In the first stage, the children are pushed into the associated parent node container, and in the second stage, the queue is filled with new potential lamination candidates.
This while loop ends as soon as the queue is empty, returning the root node, which is connected to all its descendants that are needed to construct the recursive derivative.

\centerline{
\begin{minipage}{0.87\textwidth}
	\begin{algorithm}[H]
		\begin{algorithmic}[1]
			\State{\textbf{Input:} $\boldsymbol{F}$, $\mathcal{N}$, \texttt{laminationforrest}, \texttt{startdepth}}
			\State{\textbf{Output:} \texttt{root}}
			\LineComment{\small Initialize tree with empty children array and $\xi =1.0$}
			\State{\texttt{root = Tree($\boldsymbol{F}$,1.0,startdepth+1,[])}}
			\State{initialize \texttt{queue}}
			\If{$\boldsymbol{F} \in \mathcal{N}$}
			\State{\texttt{laminate = getlaminate(}$\boldsymbol{F}$,\texttt{laminationforrest,startdepth)}}
			\State{\texttt{push(queue, (laminate,root))}}
			\Else
            \State{\texttt{points,weights = decompose(}$\boldsymbol{F},\mathcal{N}$\texttt{)}}
            \For{\texttt{(weight,point)} $\in$ \texttt{(weights,points)}}
            \State{\texttt{depth = highestlaminateorder(point,laminationforrest)}}
            \State{\texttt{candidate = Tree(point,weight,depth,[])}}
            \State{\texttt{push(queue,(candidate,root))}}
            \EndFor
			\EndIf
    			\While{\texttt{queue }$\neq \emptyset$}
    			    \State{\texttt{(lc,parent) = pop(queue)}}
    			    \If{\texttt{lc is a laminate}}
                    \State{$\xi = (\|\texttt{parent.}\boldsymbol{F} - \texttt{lc.}\boldsymbol{F}^-\|_1)/(\|\texttt{lc.}\boldsymbol{F}^+ - \texttt{lc.}\boldsymbol{F}^-\|_1)$}
                    \State{\texttt{push(parent.children,Tree(node.$\boldsymbol{F}^-,(1-\xi),\texttt{lc.}k,$[]))}}
                    \State{\texttt{push(parent.children,Tree(node.$\boldsymbol{F}^+,\xi,\texttt{lc.}k,$[]))}}
    			    \Else
                    \State{\texttt{push(parent.children,lc)}}
    			    \EndIf
    			    \For{\texttt{child $\in$ parent.children}}
                        \State{\texttt{depth $=$ child.$k - 1$}}
                        \If{\texttt{child.}$\boldsymbol{F} \in \mathcal{N}$}
                			\State{\texttt{laminate = getlaminate(child.}$\boldsymbol{F}$,\texttt{laminationforrest,depth)}}
                			\State{\texttt{push(queue, (laminate,child))}}
                        \Else
                            \State{\texttt{points,weights = decompose(child.}$\boldsymbol{F},\mathcal{N}$\texttt{)}}
                            \For{\texttt{(weight,point)} $\in$ \texttt{(weights,points)}}
                                \State{\texttt{candidate = Tree(point,weight,depth+1,[])}}
                                \State{\texttt{push(queue,(candidate,child))}}
                            \EndFor
                        \EndIf
    			    \EndFor
    			\EndWhile{}
			\State \textbf{return} \texttt{root}
		\end{algorithmic}
		\caption{\texttt{buildtree($\boldsymbol{F}, \mathcal{N}$, laminationforrest, startdepth)}}
		\label{alg:tree}
	\end{algorithm}
\end{minipage}
}~\\
With the tree at hand, the derivative can be recursively evaluated according to \eqref{eq:recursivederivative}.
A pseudocode of this procedure is given in Algorithm \ref{alg:derivative}.
The constructed tree is traversed until a leaf node is found and, in this case, the original non-convex function and its derivatives can be evaluated.
Afterwards, the values are returned and multiplied by the convex combination or weighting coefficient, respectively.

\centerline{
\begin{minipage}{0.87\textwidth}
	\begin{algorithm}[H]
		\begin{algorithmic}[1]
			\State{\textbf{Input:} \texttt{tree, material, history}}
			\State{\textbf{Output:} $W,\boldsymbol{P}, \mathbb{A}$}
			\State{$W = 0, \boldsymbol{P} = \boldsymbol{0}, \mathbb{A} = \mathbbold{0}$}
			\State{$n= 1$}
			\If{\texttt{tree.children} $=\emptyset$}\Comment{\texttt{tree} is a leafnode}
			\State{$W \pluseq W(\texttt{tree.}\boldsymbol{F})$,
    			   $\boldsymbol{P} \pluseq \partial W(\texttt{tree.}\boldsymbol{F})$,
			       $\mathbb{A} \pluseq \partial^2 W(\texttt{tree.}\boldsymbol{F})$}
			\Else
			\For{\texttt{child} $\in$ \texttt{tree.children}}
    			\State{$W_c, \boldsymbol{P}_c, \mathbb{A}_c =$ \texttt{eval(child,material,history)}} \Comment{recursion call}
			\State{$W \pluseq \texttt{child.}\xi \, W_c$,
    			   $\boldsymbol{P} \pluseq \texttt{child.}\xi \, \boldsymbol{P}_c$,
			       $\mathbb{A} \pluseq \texttt{child.}\xi \, \mathbb{A}_c$}
			\EndFor{}
			\EndIf
			\State \textbf{return} $W, \boldsymbol{P}, \mathbb{A}$
		\end{algorithmic}
		\caption{\texttt{eval(tree,material,history)}: Recursive function and derivative value construction}
		\label{alg:derivative}
	\end{algorithm}
\end{minipage}
}~\\

An alternative approach for the construction of derivative information is to consider a representative subdifferential of the constructed multidimensional linear interpolation of the rank-one convex envelope, whenever an entry of the deformation gradient lies on a cell boundary.
Multiple subdifferentials in the approximate neighborhood need to be evaluated and averaged.
However, at points where change of signs in the subdifferential occurs as it is the case at the transition from compression to tension, a special treatment is required.
Since the compression and tension regime are not of the same size, symmetry is not given and, thus, the subdifferential average is distorted.
Hence, the subdifferential is set to zero in this case.

\subsection{Overall procedure}
For a given boundary value problem, in each pseudo-time iterate a nonlinear solver is utilized where the rank-one convexification is performed in each quadrature point. 
The construction of the derivatives was already outlined and hence a steepest descent method or a Newton scheme can be applied with a line search or Armijo–Goldstein stepsize criterion.
For steepest descent and Armijo–Goldstein, see \cite[Algorithm 9.1]{Bar:2015:nmn}; for Newton, see \cite[Algorithm 4.2]{Bar:2015:nmn}.
The rank-one convexification is performed with a direction set \eqref{eq:redrankonedir} or \eqref{eq:bartelsdirections} according to the optimization problem \eqref{eq:minglobiter}.
From a computational point of view, the direction set \(\mathcal{R}\) is obtained by creating all vectors \(\boldsymbol{a}, \boldsymbol{b}\) whose outer products characterize the direction set.

\subsection{Implementation details}
Implementation details play a crucial role to get into a range of computational efficiency where a concurrent computation of the rank-one convex envelope is feasible.
Thus, this subsection discusses implementation decisions.
Our implementation in Julia \cite{BezKarShaEde:2012:jfd} enables a significant gain in computational efficiency by parallelization of the one-dimensional convexifications.
Each deformation gradient and, thus, the convolution over the deformation gradient grid is parallelized.
A crucial step for the parallelization is the use of \textit{lazy datastructures}, cf.~\cite{Oka:1998:le}.
Here, the notion of \textit{lazy datastructures} refers to a datastructure whose underlying data generation is postponed to the very moment it is queried.
In particular, this means for the deformation gradient grid that each grid point is computed ad hoc and not saved in a large multidimensional array.
With an explicit data structure, which buffers the full deformation grid in an array, we observed significant performance degradation.
Another import part is the parallelization strategy, which is like the concept of \textit{workstreams}, cf.~\cite{TurKroBan:2016:wdp}.
Each deformation gradient point in the deformation gradient grid formulates a task and for each thread an associated buffer is assigned, such that the allocation is thread-wise minimal.
The only difference to the \textit{workstreams} approach is that no coloring is required since the convexification grid points can be treated independently.
The tensorial representation of the deformation gradient, stresses and tangent moduli is realized by the continuum mechanics tensors library Tensors.jl \cite{CarEkr:2019:tjt}.
The multi-dimensional linear interpolation of Interpolations.jl \cite{LycHolKitCon:2022:ij} is used for the construction of $W^k_{\delta,r}$ and boundary value problems are assembled with the finite element toolbox Ferrite.jl \cite{CarEkrCon:2021:fj}.
The code for the numerical rank-one convexification can be found in the following git repository \texttt{https://github.com/koehlerson/NumericalRelaxation.jl}.

%

\section{Numerical experiments}\label{sec:examples}
This section demonstrates the functionality of the proposed rank-one convexification algorithm.
First, the convexification procedure is numerically analyzed when applied to sample energy densities. 
Then, the efficiency of the proposed algorithmic extensions is illustrated in a series of numerical experiments (Subsections \ref{sec:examples:convergence} -- \ref{sec:examples:scaling}).
In the Subsections \ref{sec:examples:uniaxial} -- \ref{sec:examples:triaxial} we proceed with the examination of several boundary value problems in terms of mesh insensitivity and their ability to capture strain softening. 
Finally, we conclude with a microstructural interpretation of the relaxation process in Subsection \ref{sec:examples:microstructure}.

The presented experiments are based on the effective strain energy densities \(\psi^{0}\) of the St.~Venant--Kirchhoff (STVK) model \eqref{eq:st-VK} and the Neo-Hooke (NH) model \eqref{eq:neo-Hookean}. 
For the initial examples in two spatial dimensions, we consider a mesh \(\mathcal{N}\) that is parameterized by \(F_{ij}^{\text{min}} \leq F_{ij}\leq F_{ij}^{\text{max}}\) and \(\delta_{ij}\) for \(i, j = 1,\ldots, d\).
Hence, every matrix component is discretized individually as discussed in Section \ref{sec:adaptive}. 
We use the same discretization for both diagonal elements represented by the parameters \(F_{ii}^{\text{min}}, F_{ii}^{\text{max}}\) and \(\delta_{ii}\).
A different discretization represented by the parameters \(F_{ij}^{\text{min}}, F_{ij}^{\text{max}}\) and \(\delta_{ij}\) for \(i \neq j\) is utilized for both off-diagonal entries.

As a stopping criterion for the global loop of the convexification procedure, we apply \(\max_{\boldsymbol{F} \in \mathcal{N}} |W^{k + 1}_{\delta, r} - W^{k}_{\delta, r}|_\infty \leq 10^{-4}\) while at the same time guaranteeing that a total of \(k_{\text{max}}=20\) iterations is not exceeded.
The number \(k_{\text{max}}\) corresponds to the maximal lamination depth of the microstructure evolution.

\subsection{Convergence for the different direction sets} \label{sec:examples:convergence}
First we analyze the distance between two consecutive iterates to illustrate the convergence behavior of the algorithm.
We compare the obtained approximations of the rank-one convex envelopes when using the reduced rank-one set $\mathcal{R}^1_1$ and the full direction set $\mathcal{R}^1_{\delta,r}$, the latter one can be interpreted as reference solution.
For this purpose, a material point study for the incremental potential \(W\) (as in \eqref{eq:ISP}) is carried out for both, an NH and an STVK effective strain energy density \(\psi^{0}\).
The material parameters and the convexification grid parameters used are listed in Table~\ref{tab:2Dconvexification}. 
Note that we used the same discretization stepsize \(\delta_{ij}\) for all matrix components.
\begin{table}
	\smaller
	\begin{tabular}{|c|c|c|c|c|c|c|c|c|c|c|c|}
		\hline
		$\psi^0$ & $\lambda$ & $\mu$ & $D_0$ & $D_\infty$ & $\beta_k$ & $\delta_{ij}$ & $F_{ii}^{\text{min}}$ & $F_{ii}^{\text{max}}$ & $F_{ij}^{\text{min}} \, \forall i\neq j$ & $F_{ij}^{\text{max}} \, \forall i\neq j$ & $|\mathcal{N}|$\\
		\hline
		NH & 0.5 & 1.0 & 0.3 & 0.9 & 0.06 & 0.15 & 0.1 & 3.4 & -2.55 & 2.55 & 648025\\
		\hline
		STVK & 0.5 & 1.0 & 0.3 & 0.9 & 0.07 & 0.1 & 0.1 & 2.0 & -2.0 & 2.0 & 672400\\
		\hline
	\end{tabular}
    \caption{Material and convexification parameters for the $2 \times 2=4$ fully resolved deformation grid material point study.}
    \label{tab:2Dconvexification}
\end{table}
The tables presented in Figure \ref{fig:convergence} show the convergence of the algorithm for an incremental stress potential \(W\) as in \eqref{eq:ISP} based on the given parameter set in Table \ref{tab:2Dconvexification}.
In the case of the STVK model, 1000 was set as the starting value which was chosen instead of setting it to $\infty$.
The maximal decrease 
\begin{align} \label{eq:maxdiff}
	\max_{\boldsymbol{F} \in \mathcal{N}} \, W^{k}_{\delta, r}(\boldsymbol{F}) - W^{k+1}_{\delta, r}(\boldsymbol{F})
\end{align}
in iteration \(k\) is compared for both, the full direction set \eqref{eq:bartelsdirections} and the reduced direction set \eqref{eq:redrankonedir}.
The results show that the reduced set of directions is enough and information of decrease in function values of the mesh points is sufficiently propagated by the smaller set of directions.
The observed increase in iterations and hence slower convergence of the reduced directions is manageable since the overall costs massively outperform the full rank-one set.
The number of elements in the set $\mathcal{R}_1^1$ is sixteen, which is significantly lower than the number of elements in the $\mathcal{R}_{\delta,r}^1$ set, which consists of 93925 and 122199 elements for the NH and SVKT parameters given in Table \ref{tab:2Dconvexification}, respectively.
The quality of the converged hull is compared in the next subsection.
\begin{figure} \label{fig:convergenceNHSTVK}
\begin{subfigure}[b]{0.5\textwidth}
	\centering 
	\begin{tabular}{c|c c} \label{fig:convergenceSTVK} 
		$k$ & $\mathcal{R}_{\delta,r}^1$ & $\mathcal{R}^1_1$ \\
		\hline
		1  & 1000.00 & 1000.00 \\
		2  & 1000.04 & 1000.04 \\
		3  & 0.18478 & 0.18478 \\
		4  & 0.02101 & 0.07939 \\
		5  & 0.00278 & 0.01625 \\
		6  & 0.00105 & 0.00237 \\
		7  & 0.00062 & 0.00132 \\
		8  & 0.00043 & 0.00075 \\
		9  & 0.00013 & 0.00045 \\
		10 & -       & 0.00037 \\
		11 & -       & 0.00014 \\
		12 & -       & - \\
		13 & -       & - \\
		14 & -       & - \\
		15 & -       & - \\
	\end{tabular}
	\caption{}
\end{subfigure}
\hfill
\begin{subfigure}[b]{0.49\textwidth}
	\centering
	\begin{tabular}{c|c c} \label{fig:convergenceNH} 
		$k$ & $\mathcal{R}_{\delta,r}^1$ & $\mathcal{R}^1_1$ \\
		\hline
		1  & 2.84696 & 2.84696 \\
		2  & 1.68770 & 1.62022 \\
		3  & 1.14215 & 1.14215 \\
		4  & 0.01503 & 0.09311 \\
		5  & 0.00453 & 0.01674 \\
		6  & 0.00174 & 0.00401 \\
		7  & 0.00072 & 0.00182 \\
		8  & 0.00044 & 0.00084 \\
		9  & 0.00026 & 0.00056 \\
		10 & 0.00012 & 0.00042 \\
		11 & -       & 0.00031 \\
		12 & -       & 0.00027 \\
		13 & -       & 0.00020 \\
		14 & -       & 0.00013 \\
		15 & -       & 0.00011 \\
	\end{tabular}
	\caption{}
\end{subfigure}
\caption{Distance of consecutive iterates \eqref{eq:maxdiff} for the models based on the strain energy densities \(\psi^{0}\) for the STVK (left) and NH (right) model, stopping tolerance \(10^{-4}\).}
\label{fig:convergence}
\end{figure}

\subsection{Comparison of reduced and full rank-one direction sets}
\begin{figure}
\begin{subfigure}[b]{0.32\textwidth}
	\centering
	\ifthenelse{\boolean{professormode}}
	{\includegraphics{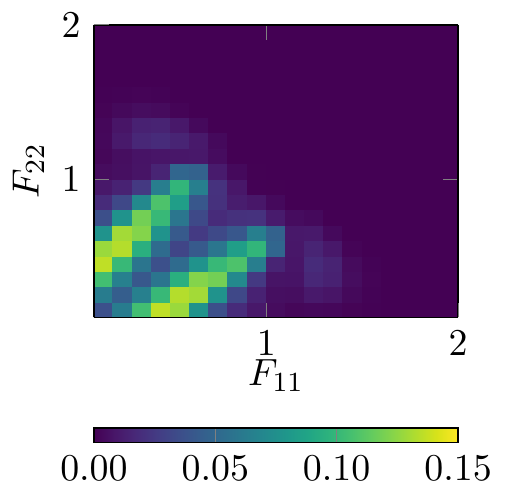}}
	{\input{figures/tikz/svk-diff-biaxial-plane}}
\end{subfigure}
\hfill
\begin{subfigure}[b]{0.32\textwidth}
	\centering
	\ifthenelse{\boolean{professormode}}
	{\includegraphics{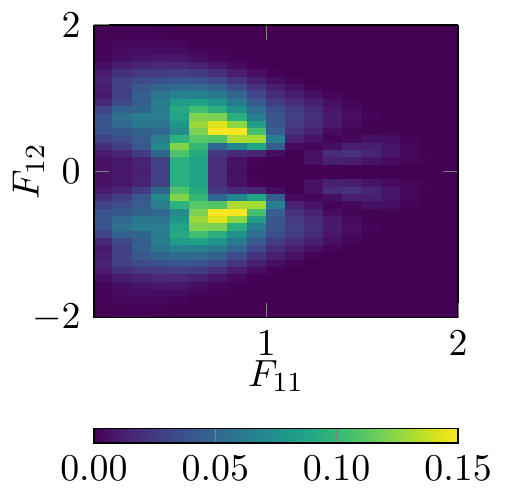}}
	{\input{figures/tikz/svk-diff-diagonal-offdiagonal-plane}}
\end{subfigure}
\hfill
\begin{subfigure}[b]{0.32\textwidth}
	\centering
	\ifthenelse{\boolean{professormode}}
	{\includegraphics{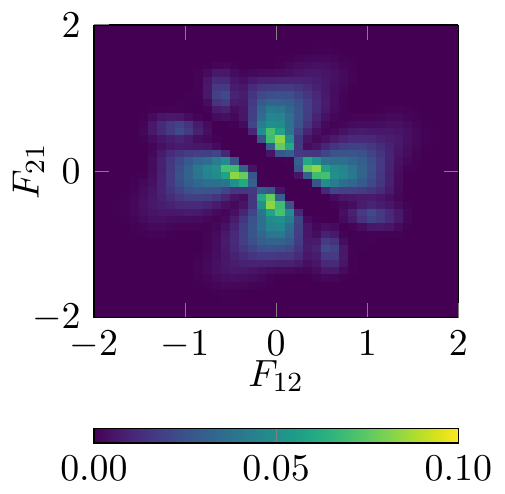}}
	{\input{figures/tikz/svk-diff-offdiagonal-plane}}
\end{subfigure}
\caption{Relative error $\mathcal{E}^{\text{rel}}$ for $W_{\delta,r,\mathcal{R}^1_{\delta,r}}^9$ and $W_{\delta,r,\mathcal{R}^1_{1}}^{11}$ obtained by the direction sets $\mathcal{R}^1_1$ and $\mathcal{R}_{\delta,r}^1$ based on the STVK model for $\psi^0$.}
\label{fig:svkdiff}
\end{figure}

\begin{figure}
\begin{subfigure}[b]{0.32\textwidth}
	\centering
	\ifthenelse{\boolean{professormode}}
	{\includegraphics{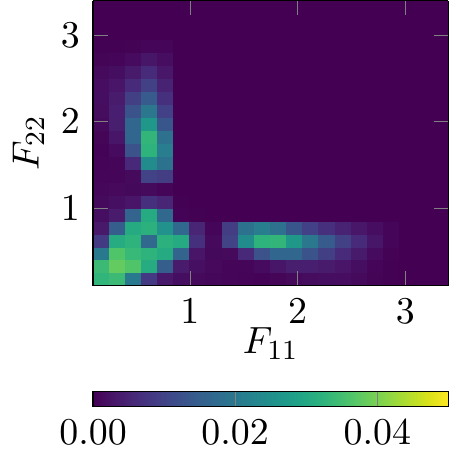}}
	{\input{figures/tikz/neohooke-diff-biaxial-plane}}
\end{subfigure}
\hfill
\begin{subfigure}[b]{0.32\textwidth}
	\centering
	\ifthenelse{\boolean{professormode}}
	{\includegraphics{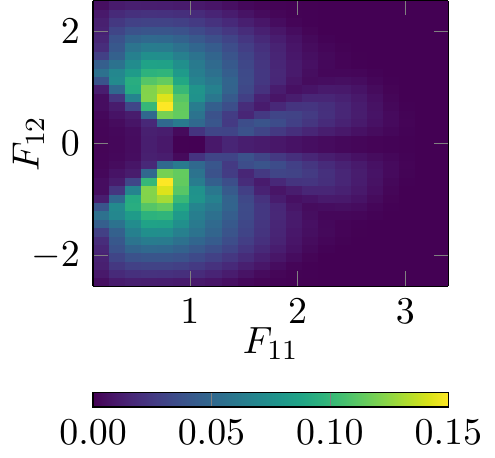}}
	{\input{figures/tikz/neohooke-diff-diagonal-offdiagonal-plane}}
\end{subfigure}
\hfill
\begin{subfigure}[b]{0.32\textwidth}
	\centering
	\ifthenelse{\boolean{professormode}}
	{\includegraphics{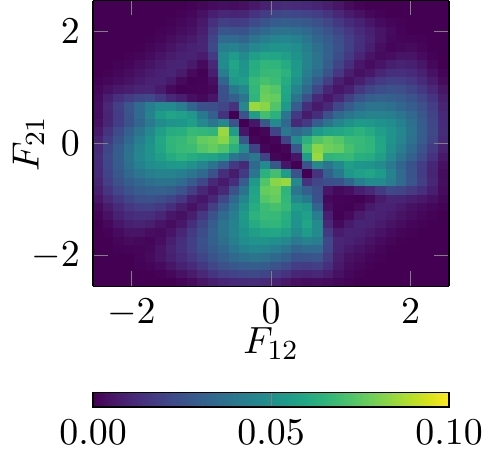}}
	{\input{figures/tikz/neohooke-diff-offdiagonal-plane}}
\end{subfigure}
\caption{Relative error $\mathcal{E}^{\text{rel}}$ for $W_{\delta,r,\mathcal{R}^1_{\delta,r}}^{10}$ and $W_{\delta,r,\mathcal{R}^1_{1}}^{15}$ obtained by the direction sets $\mathcal{R}^1_1$ and $\mathcal{R}_{\delta,r}^1$ based on the NH model for $\psi^0$.}
\label{fig:neohookediff}
\end{figure}

Figures \ref{fig:svkdiff} and \ref{fig:neohookediff} show the relative error of the approximations of the two direction sets, i.e., 
\begin{align*}
	\mathcal{E}^{\text{rel}}(\boldsymbol{F}) = \frac{\Big\vert\, W_{\delta,r,\mathcal{R}^1_{\delta,r}}^{\hat{k}}(\boldsymbol{F}) - W_{\delta,r,\mathcal{R}^1_1}^k(\boldsymbol{F})\,\Big\vert}{\gamma + \Big\vert\, W_{\delta,r,\mathcal{R}^1_{\delta,r}}^{\hat{k}}(\boldsymbol{F})\,\Big\vert}
\end{align*} 
for \(\boldsymbol{F} \in \mathcal{N}\).
Here, \(\hat{k}\) corresponds to the converged iteration index of the reference approximation computed by the full rank-one direction set and \(k\) corresponds to the iteration for the converged approximation computed by the reduced direction set.
The stablization parameter $\gamma$ is introduced, since both functions exhibit a value of zero multiple times and, thus, the division by zero produces undefined values. The stabilization parameter has been set to $\gamma=10^{-8}$.

For all plotted planes in Figures \ref{fig:svkdiff} and \ref{fig:neohookediff} the fixed axes correspond to initial, undeformed configuration deformation values $\boldsymbol{F}_0=\boldsymbol{I}$.
In other words, all plots show slices of the mesh hypercuboid that contain the identity matrix, i.e. the point \(\boldsymbol{I}\).
Compared to the absolute function values of \(W^{k}_{\delta, r}\), the difference in the two possible approximations is quite small from a quantitative perspective.
Hence, replacing the full rank-one direction set \(\mathcal{R}_{\delta, r}^1\) by \(\mathcal{R}_1^1\) seems reasonable in this representative model problem.

Notice that due to the asymmetry of the compression (\(0 \leq {F}_{ii} \leq 1\)) and tension (\(1 \leq {F}_{ii}\)) regime in the \(\R^{d \times d}\) space, the equidistant discretization favors the tension regime in terms of error distribution.
This is due to the smaller size of the compression region and therefore coarser approximation, while the same discretization step size in the tension range is able to capture the behavior of the incremental stress potential slightly better because of the slower function growth for \(F_{ii} \to \infty\). 
For an overall better approximation in the compression zone, one should prefer a finer discretization in the deformation gradient space instead of taking a larger rank-one direction set. 
The information of 'function lowering' due to convexification seems to be propagated by the reduced direction sufficiently (requiring a higher number of global iterations) and a refinement in terms of directions seems to raise no substantial gain.
Hence, with regard to refinement at the cost of higher workload, the spatial resolution of the deformation gradient should be improved instead of increasing the number of rank-one directions.

\subsection{Lamination matrix} 
\begin{figure}
\begin{subfigure}[b]{0.32\textwidth}
	\centering
	\ifthenelse{\boolean{professormode}}
	{\includegraphics{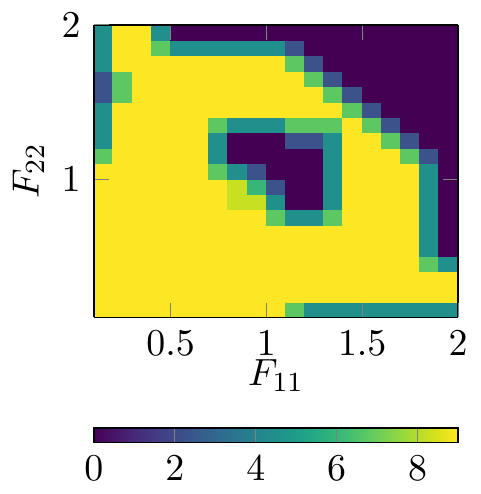}}
	{\input{figures/tikz/lamination_matrix-svk_bartels-biaxial}}
	\label{fig:lamination_matrix-svk_bartels-biaxial0}
\end{subfigure}
\hfill
\begin{subfigure}[b]{0.32\textwidth}
	\centering
	\ifthenelse{\boolean{professormode}}
	{\includegraphics{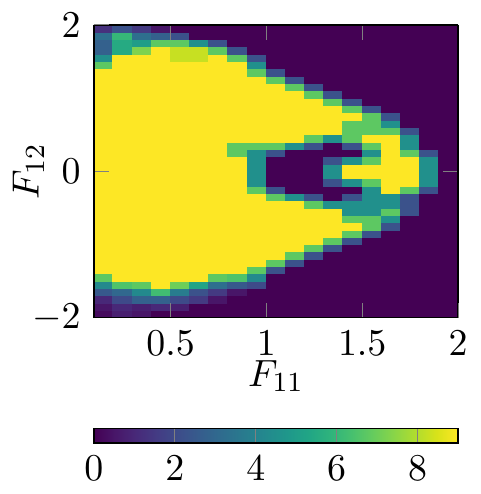}}
	{\input{figures/tikz/lamination_matrix-svk_bartels-diagonal-offdiagonal}}
	\label{fig:lamination_matrix-svk_bartels-diagonal-offdiagonal0}
\end{subfigure}
\hfill
\begin{subfigure}[b]{0.32\textwidth}
	\centering
	\ifthenelse{\boolean{professormode}}
	{\includegraphics{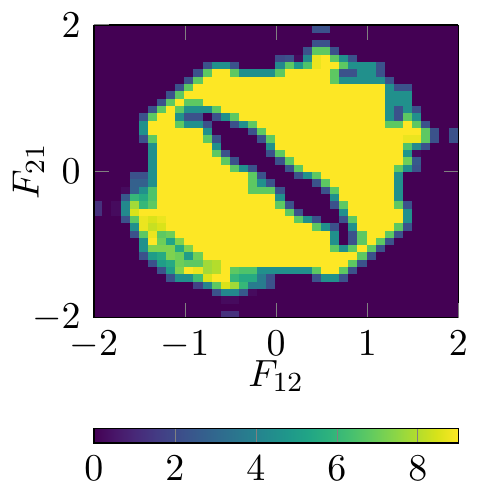}}
	{\input{figures/tikz/lamination_matrix-svk_bartels-offdiagonal}}
	\label{fig:lamination_matrix-svk_bartels-offdiagonal0}
\end{subfigure}
\begin{subfigure}[b]{0.32\textwidth}
	\centering
	\ifthenelse{\boolean{professormode}}
	{\includegraphics{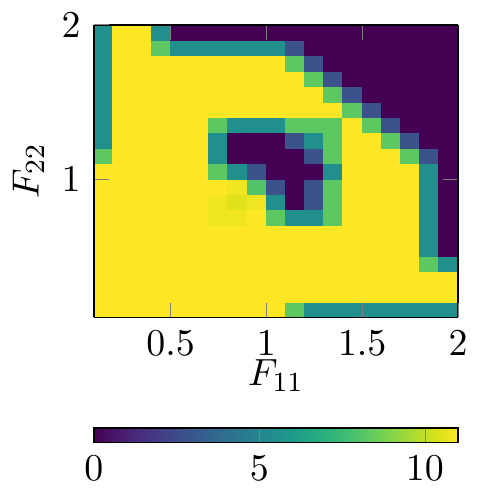}}
	{\input{figures/tikz/lamination_matrix-svk_principal-biaxial}}
\end{subfigure}
\hfill
\begin{subfigure}[b]{0.32\textwidth}
	\centering
	\ifthenelse{\boolean{professormode}}
	{\includegraphics{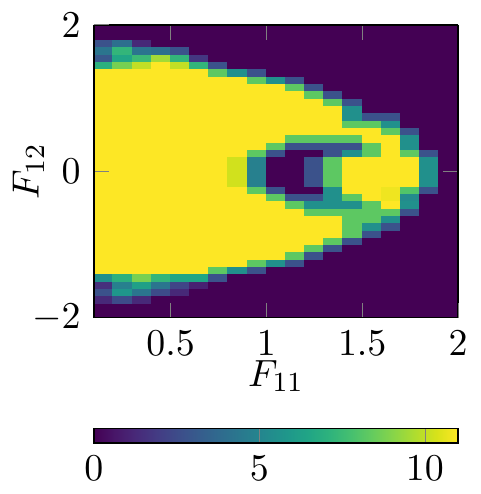}}
	{\input{figures/tikz/lamination_matrix-svk_principal-diagonal-offdiagonal}}
\end{subfigure}
\hfill
\begin{subfigure}[b]{0.32\textwidth}
	\centering
	\ifthenelse{\boolean{professormode}}
	{\includegraphics{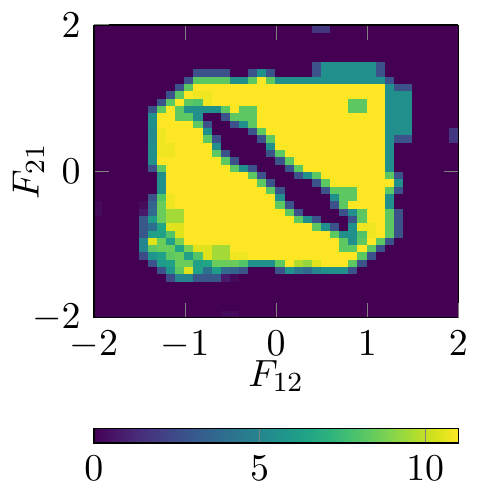}}
	{\input{figures/tikz/lamination_matrix-svk_principal-offdiagonal}}
\end{subfigure}
\caption{Matrix of lamination order for St.~Venant--Kirchhoff $\psi^0$.
The color corresponds to the iteration $k$ for which the last time a laminate was built.
(top) shows the obtained lamination matrix for \(\mathcal{R}_{\delta,r}^1\) and (bottom) in the case of \(\mathcal{R}_1^1\).
Note that the convex regime, i.e. the area, where either a fully dissipated configuration (\(D=D_\infty\)) is reached or ellipticity is not lost yet (\(\boldsymbol{F}\approx\boldsymbol{I}\)), does not need lamination.}
\label{fig:lamination-svk}
\end{figure}

\begin{figure}
\begin{subfigure}[b]{0.32\textwidth}
	\centering
	\ifthenelse{\boolean{professormode}}
	{\includegraphics{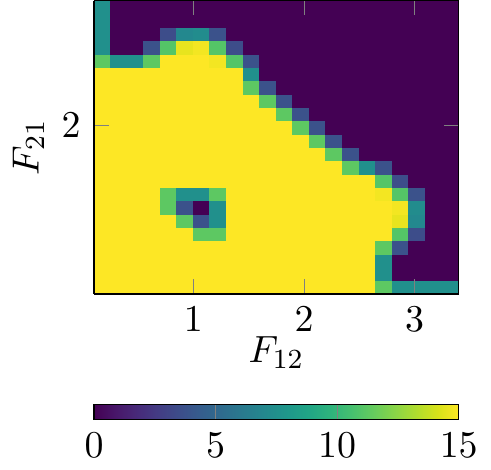}}
	{\input{figures/tikz/lamination_matrix-neohooke_principal-biaxial}}
\end{subfigure}
\hfill
\begin{subfigure}[b]{0.32\textwidth}
	\centering
	\ifthenelse{\boolean{professormode}}
	{\includegraphics{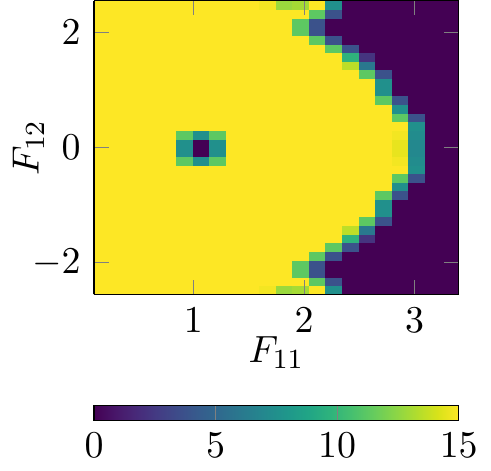}}
	{\input{figures/tikz/lamination_matrix-neohooke_principal-diagonal-offdiagonal}}
\end{subfigure}
\hfill
\begin{subfigure}[b]{0.32\textwidth}
	\centering
	\ifthenelse{\boolean{professormode}}
	{\includegraphics{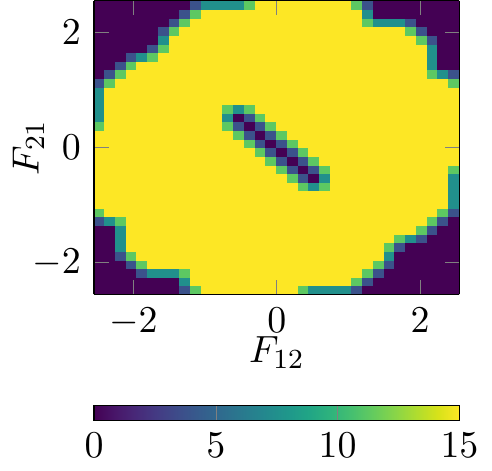}}
	{\input{figures/tikz/lamination_matrix-neohooke_principal-offdiagonal}}
\end{subfigure}
\begin{subfigure}[b]{0.32\textwidth}
	\centering
	\ifthenelse{\boolean{professormode}}
	{\includegraphics{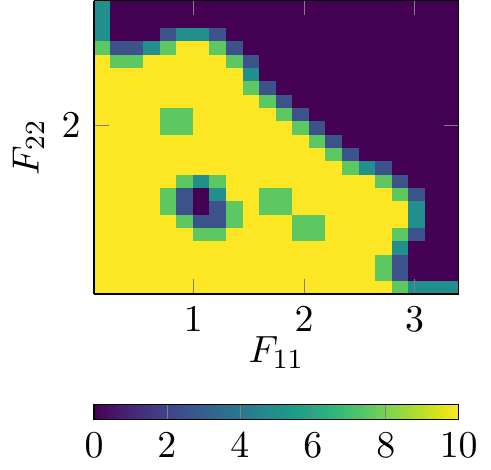}}
	{\input{figures/tikz/lamination_matrix-neohooke_bartels-biaxial-plane}}
\end{subfigure}
\hfill
\begin{subfigure}[b]{0.32\textwidth}
	\centering
	\ifthenelse{\boolean{professormode}}
	{\includegraphics{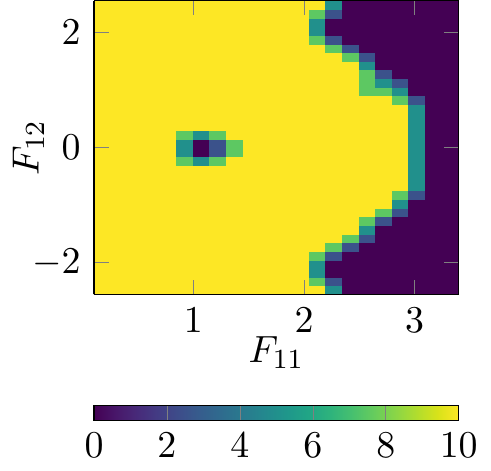}}
	{\input{figures/tikz/lamination_matrix-neohooke_bartels-diagonal-offdiagonal-plane}}
\end{subfigure}
\hfill
\begin{subfigure}[b]{0.32\textwidth}
	\centering
	\ifthenelse{\boolean{professormode}}
	{\includegraphics{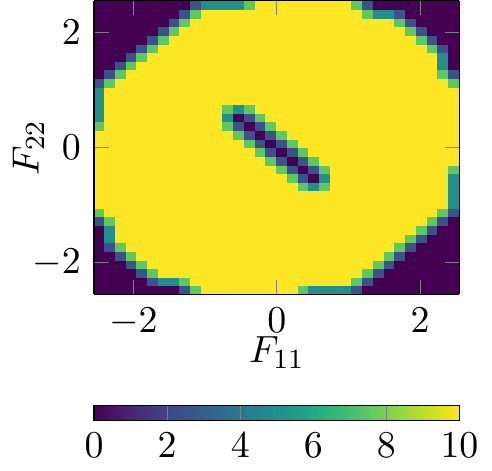}}
	{\input{figures/tikz/lamination_matrix-neohooke_bartels-offdiagonal-plane}}
\end{subfigure}
\caption{Lamination matrix for Neo-Hooke effective strain energy density.
Again, the qualitative behavior is similar between the (top): \(\mathcal{R}_{\delta,r}\) direction set and (bottom): \(\mathcal{R}^1_1\) approach}
\label{fig:lamination-neohooke}
\end{figure}

The Figures \ref{fig:lamination-svk} and \ref{fig:lamination-neohooke} show slices of the computed states corresponding to the last row of the tables in Figure \ref{fig:convergence} for the STVK and NH model, respectively. 
The pictures show lamination values on slices of the mesh hypercuboid through the point \(\boldsymbol{I}\) as described in the previous subsection.
For two given matrix component values, the lamination order for the full \eqref{eq:bartelsdirections} and reduced directions \eqref{eq:redrankonedir} are illustrated. 
Again, the higher lamination order of the reduced directions is observable while the qualitative behavior is similar. 
Also the convex (lamination order \(0\)) regime around the identity matrix \(\boldsymbol{I}\) is visible. This illustration also motivates the replacement of \(\mathcal{R}_{\delta, r}^{1}\) by \(\mathcal{R}_{1}^{1}\).

\subsection{Strong scaling of the rank-one convexification} \label{sec:examples:scaling}
\begin{figure}
	\tikzsetfigurename{strong-scaling-convexification-neohooke}
\begin{tikzpicture}
\begin{axis}[ymode=log,
             xmode=log,
             axis background/.style={{fill={white!89.803921568!black}}},
             x grid style={{white}},
             y grid style={{white}},
             xmajorgrids,
             ymajorgrids,
             ylabel={Time in (s)},
             xlabel={Number of threads},
             legend style={{at={(0.5,-0.2)},anchor=north}}]
    \addplot[color={red!60!gray}, mark={x}, mark options={scale=2}, ultra thick]
        coordinates {
            (1,1960.739)
            (2,988.585)
            (4,512.932)
            (8,270.08)
            (16,147.04)
            (32,81.52)
            (64,49.06)
        }
        ;
    \addlegendentry {Rank-One Convexification Implementation $\mathcal{R}^1_1$}
    \addplot[color={blue!60!gray}, mark={x}, mark options={scale=2}, ultra thick]
        coordinates {
            (1,86779.36)
            (2,43401.14)
            (4,22823.34)
            (8,11782.19)
            (16,6169.88)
            (32,3081.38)
            (64,1602.37)
        }
        ;
    \addlegendentry {Rank-One Convexification Implementation $\mathcal{R}^1_{\delta,r}$}
    \addplot[color={gray!80!red}, dashed, mark={x}, mark options={scale=2}, thick]
        coordinates {
            (1,1960.739)
            (2,980.3695)
            (4,490.18475)
            (8,245.092375)
            (16,122.5461875)
            (32,61.27309375)
            (64,30.636546875)
        }
        ;
    \addlegendentry {Perfect Scaling,$\mathcal{R}^1_1$}
    \addplot[color={gray!80!blue}, dashed, mark={x}, mark options={scale=2}, thick]
        coordinates {
            (1,86779.36)
            (2,43389.68)
            (4,21694.84)
            (8,10847.42)
            (16,5423.71)
            (32,2711.855)
            (64,1355.9275)
        }
        ;
    \addlegendentry {Perfect Scaling,$\mathcal{R}^1_{\delta,r}$}
\end{axis}
\end{tikzpicture}
	\caption{Strong scaling of Neo-Hooke example with a convexification mesh consisting of 43681 points on a cluster node with an Intel Xeon Phi 7210 processor.
	The scaling consists of fixing the total workload while varying the number of threads by 1, 2, 4, 8, 16, 32, and 64.}
	\label{fig:neohookescaling}
\end{figure}
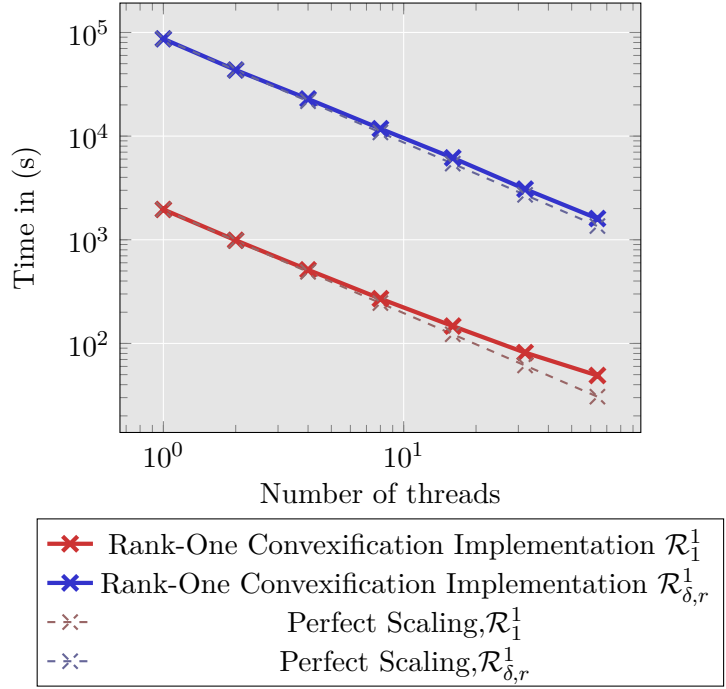

The parallelized version of the convexification algorithm is tested in a strong scaling setting on a single cluster node with an increasing number of threads with a fixed amount of total work, i.e.~convexification grid size. 
The scaling study was done without building the decomposition tree.
If the tree is needed either each thread has its own partial lamination forest which needs to be merged after the parallel part of the code or a parallel safe data structure for the lamination forest can be used.
In our implementation, each thread has its own partial lamination forest that is merged after the parallel part of the code.
The implementation's performance can be seen in Figure \ref{fig:neohookescaling} which is close to the logarithmic perfect scaling. 
Julia's garbage collector is realized by a \textit{stop the world} implementation, cf. \cite[Section 7]{NasBezPam:2019:bte}.
Therefore, the gain of having the deformation grid not buffered in the memory comes with a tradeoff.
While, initially, there is a gain in performance due to the avoided paging faults, cache misses and similar effects, the performance degrades from perfect scaling since there are more threads whose local memory needs to be cleaned in a serialized way after finishing the threaded convexification.
However, since convexification meshes scale drastically in memory, we still advocate for lazy convexification grid representation.

Until now, the focus was on the pure application of the convexification algorithm, its convergence and lamination properties using different direction sets.
In the next subsections, we test the obtained rank-one convex hull approximations in terms of mesh dependence under different loading and material conditions.

\subsection{Uniaxial mesh independence test} \label{sec:examples:uniaxial}
\begin{figure}
\centering
\begin{subfigure}[b]{0.3\textwidth}
	\centering
	\ifthenelse{\boolean{professormode}}
	{\includegraphics{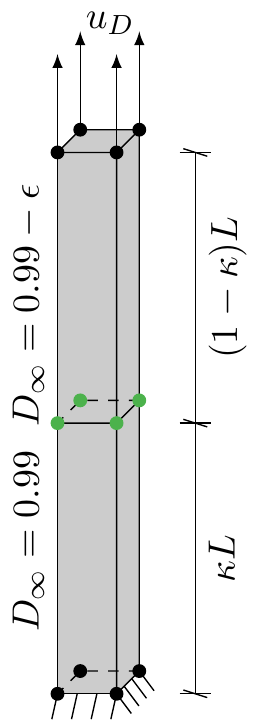}}
	{\tikzsetfigurename{perturbation-problem3D}
\begin{tikzpicture}
    \pgfmathsetmacro{\cubex}{0.6}
    \pgfmathsetmacro{\cubey}{5.5}
    \pgfmathsetmacro{\cubez}{0.6}
    \pgfmathsetmacro{\distance}{0.2}
    \pgfmathsetmacro{\shift}{0.8}
    \pgfmathsetmacro{\linelength}{0.2}
    \pgfmathsetmacro{\inclination}{0.15}
    \pgfmathsetmacro{\shiftz}{0.1}
    \pgfmathsetmacro{\nodesize}{4.0}
    \pgfmathsetmacro{\Dshift}{1.2}
    \pgfmathsetmacro{\arrowlength}{1.0}
    \draw[black,fill=gray!40] (0,0,0) -- ++(-\cubex,0,0) -- ++(0,-\cubey,0) node[midway,above,rotate=90,xshift=-\Dshift cm]{ $D_{\infty}=0.99$} node[midway,above,rotate=90,xshift=\Dshift cm]{ $D_{\infty}=0.99-\epsilon$} -- ++(\cubex,0,0) -- cycle;
    \draw[black,fill=gray!40] (0,0,0) -- ++(0,0,-\cubez) -- ++(0,-\cubey,0) -- ++(0,0,\cubez) -- cycle;
    \draw[black,fill=gray!40] (0,0,0) -- ++(-\cubex,0,0) -- ++(0,0,-\cubez) -- ++(\cubex,0,0) -- cycle;
    \draw[black] (-\cubex,-\cubey/2,0) -- (0,-\cubey/2,0) -- (0,-\cubey/2,-\cubez);

    \draw[black] (\shift*0.8,-\cubey/2,0) -- (\shift*1.2,-\cubey/2,0);
    \draw[black] (\shift*0.8,-\cubey,0) -- (\shift*1.2,-\cubey,0);
    \draw[black] (\shift*0.8,0,0) -- (\shift*1.2,0,0);

    \draw[black] (\shift*0.8,-\cubey/2,-\shiftz) -- (\shift*1.2,-\cubey/2,\shiftz);
    \draw[black] (\shift*0.8,-\cubey,-\shiftz) -- (\shift*1.2,-\cubey,\shiftz);
    \draw[black] (\shift*0.8,0,-\shiftz) -- (\shift*1.2,0,\shiftz);

    \draw[dashed] (0,-\cubey/2,-\cubez) -- (-\cubex, -\cubey/2, -\cubez) -- (-\cubex, -\cubey/2, 0);
    \draw[dashed] (0,-\cubey,-\cubez) -- (-\cubex, -\cubey, -\cubez) -- (-\cubex, -\cubey, 0);

    \foreach \x in {0,-\distance,...,-\cubex}
        \draw[black] (\x,-\cubey,0) -- (\x,-\cubey-\linelength,\inclination) ;
    \foreach \x in {0,-\distance,...,-\cubex}
        \draw[black] (0,-\cubey,\x) -- (\inclination,-\cubey-\linelength,\x) ;

    \draw[->,-latex] (-\cubex,0,-\cubez) -- (-\cubex,\arrowlength,-\cubez) node[right,xshift=-0.2em,yshift=0.2em]{$u_D$};
    \draw[->,-latex] (-\cubex,0,0) -- (-\cubex,\arrowlength,0);
    \draw[->,-latex] (0,0,0) -- (0,\arrowlength,0);
    \draw[->,-latex] (0,0,-\cubez) -- (0,\arrowlength,-\cubez);

    \draw[] (\shift,0,0) -- (\shift,-\cubey/2,0) node[midway,below,rotate=90]{ $(1-\kappa)L$} -- (\shift,-\cubey,0) node[midway,below,rotate=90]{ $\kappa L$};

    \draw  node[fill=black,circle,inner sep=0pt,minimum size=\nodesize pt] at (0,0,0) {};
    \draw  node[fill=black,circle,inner sep=0pt,minimum size=\nodesize pt] at (-\cubex,0,0) {};
    \draw  node[fill=black,circle,inner sep=0pt,minimum size=\nodesize pt] at (-\cubex,0,-\cubez) {};
    \draw  node[fill=black,circle,inner sep=0pt,minimum size=\nodesize pt] at (0,0,-\cubez) {};

    \draw  node[fill=green!40!gray,circle,inner sep=0pt,minimum size=\nodesize pt] at (0,-\cubey/2,0) {};
    \draw  node[fill=green!40!gray,circle,inner sep=0pt,minimum size=\nodesize pt] at (-\cubex,-\cubey/2,0) {};
    \draw  node[fill=green!40!gray,circle,inner sep=0pt,minimum size=\nodesize pt] at (-\cubex,-\cubey/2,-\cubez) {};
    \draw  node[fill=green!40!gray,circle,inner sep=0pt,minimum size=\nodesize pt] at (0,-\cubey/2,-\cubez) {};

    \draw  node[fill,circle,inner sep=0pt,minimum size=\nodesize pt] at (0,-\cubey,0) {};
    \draw  node[fill,circle,inner sep=0pt,minimum size=\nodesize pt] at (-\cubex,-\cubey,0) {};
    \draw  node[fill,circle,inner sep=0pt,minimum size=\nodesize pt] at (-\cubex,-\cubey,-\cubez) {};
    \draw  node[fill,circle,inner sep=0pt,minimum size=\nodesize pt] at (0,-\cubey,-\cubez) {};
\end{tikzpicture}}
	\caption{Uniaxial boundary value problem}
	\label{fig:uniaxial-bvp}
\end{subfigure}
\begin{subfigure}[b]{0.5\textwidth}
	\centering
	\ifthenelse{\boolean{professormode}}
	{\includegraphics{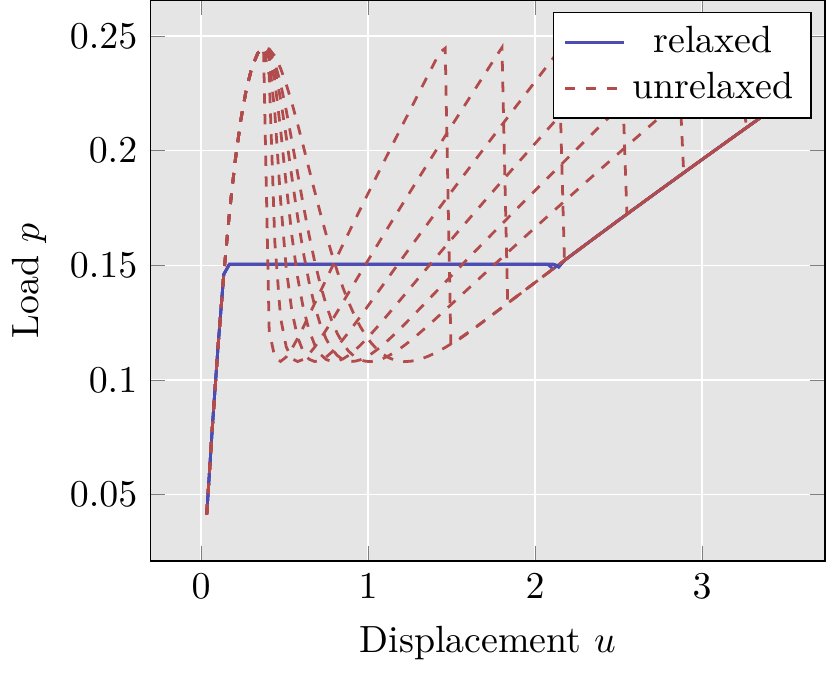}}
	{\input{figures/tikz/multid-uniaxial-neohooke}}
	\caption{Uniaxial mesh sensitivity in load displacement diagram}
	\label{fig:multid-uniaxial-neohooke}
\end{subfigure}
\label{fig:uniaxial-nh}
\caption{The same uniaxial perturbation test as in \cite{BalOrt:2012:riv,GurMie:2011:edm} for a Neo-Hooke effective strain energy density.
Due to the exhibited rank-one path of the deformation, the same convex response as in the aforementioned contributions can be seen for the relaxed model.
The unrelaxed model shows mesh dependence, where the descent method snaps back into different minima for different values of $\kappa$.}
\end{figure}

As a first mesh independence test, we consider the two element perturbation test of \cite{GurMie:2011:edm} and \cite{BalOrt:2012:riv}.
However, here bilinear quadrilateral elements are considered.
The boundary value problem is depicted in Figure \ref{fig:uniaxial-bvp} and the associated force-displacement curve in Figure \ref{fig:multid-uniaxial-neohooke}.
Within this boundary value problem, a pure uniaxial deformation is exhibited. Therefore, only the diagonal entries of the deformation gradient are resolved by the convexification grid.
In order to introduce inhomogeneity to the problem, the asymptotic damage limit $D_\infty$ in one of the two finite elements is perturbed by a small, physically meaningless value $\epsilon=10^{-5}$, such that \(D_{\infty}\) is replaced by \(D_{\infty}-\epsilon\). 
The total length $L=1$ is fixed; however, the individual element size of both elements is varied by the parameter \(\kappa\).
The derivatives were constructed with the subdifferential approach in this case.
Then, only the first variation of the potential is known, and therefore, the descent method of \cite[Algorithm 9.1]{Bar:2015:nmn} with Armijo--Goldstein based linesearch is used.
The parameters of the Armijo--Goldstein linesearch were set to $\alpha=0.5$ and \(\hat{\mu} = 0.01\), respectively.
As material parameters, the following settings were chosen: The Neo-Hookean effective strain energy density $\psi^0_{\text{NH}}$ is used with $\lambda=0.5,\mu=1.0, D_0=0.3, D_\infty=0.9$ and the convexification grid was spanned with a stepsize of 0.15 from 1.0 to 3.4 for both diagonal entries of the deformation gradient.
From the force--displacement curve, an analogous behaviour as in \cite{BalOrt:2012:riv} can be seen.
As soon as the non-convex regime is entered, the relaxed model is activated and a constant stress response is observed.
Since the deformation path can be parameterized by a rank-one path, the approximated rank-one convex envelope is convex along the path, thus the constant stresses.
Furthermore, a mesh-independent response can be observed since different discretizations $\kappa$ do not lead to a different response. This contrasts the unrelaxed basic model, which shows heavy mesh sensitivity.

\subsection{Biaxial mesh independence test} \label{sec:examples:biaxial}
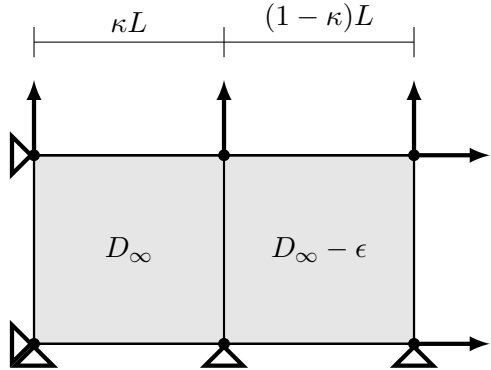
\begin{figure}
	\centering
	 \ifthenelse{\boolean{professormode}}
	 {\includegraphics{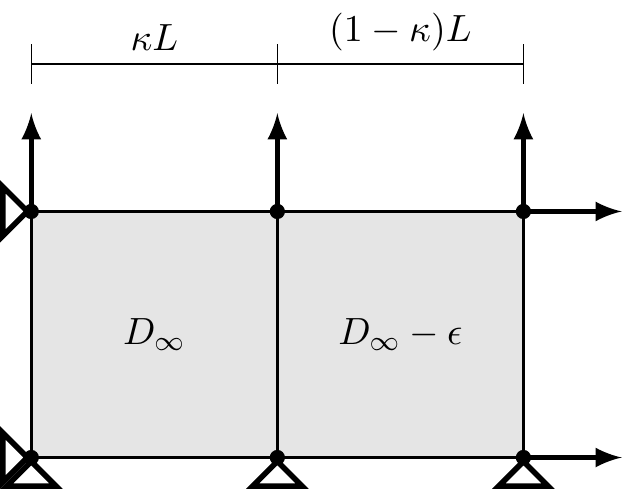}}
	 {\tikzsetfigurename{biaxial}
\begin{tikzpicture}
       \pgfmathsetmacro\sizele{5.0}
       \pgfmathsetmacro\nodesize{2.0}
       \pgfmathsetmacro\trianglesize{0.25}
       \pgfmathsetmacro\arrowsize{1.0}
       \pgfmathsetmacro\stepsize{\sizele*0.5}
       \draw[step=\stepsize,thick,fill=gray!20] (0,0) grid (\sizele,\stepsize) rectangle (0,0);
       \draw[fill=black] (0,0) circle (\nodesize pt);
       \draw[fill=black] (0,\stepsize) circle (\nodesize pt);
       \draw[fill=black] (\stepsize,\stepsize) circle (\nodesize pt);
       \draw[fill=black] (\stepsize,0) circle (\nodesize pt);
       \draw[fill=black] (\stepsize*2,\stepsize) circle (\nodesize pt);
       \draw[fill=black] (\stepsize*2,0) circle (\nodesize pt);
       \node[isosceles triangle, isosceles triangle apex angle=90,draw, inner sep=0pt, rotate=90, minimum size =\trianglesize cm,anchor=apex,ultra thick] at (0,0){};
       \node[isosceles triangle, isosceles triangle apex angle=90,draw, inner sep=0pt, rotate=0,  minimum size =\trianglesize cm,anchor=apex,ultra thick] at (0,0){};
       \node[isosceles triangle, isosceles triangle apex angle=90,draw, inner sep=0pt, rotate=0,  minimum size =\trianglesize cm,anchor=apex,ultra thick] at (0,\stepsize){};
       \node[isosceles triangle, isosceles triangle apex angle=90,draw, inner sep=0pt, rotate=90, minimum size =\trianglesize cm,anchor=apex,ultra thick] at (\stepsize,0){};
       \node[isosceles triangle, isosceles triangle apex angle=90,draw, inner sep=0pt, rotate=90, minimum size =\trianglesize cm,anchor=apex,ultra thick] at (\stepsize*2,0){};
       \draw[->,-latex,ultra thick] (\stepsize*2,0) --++ (\arrowsize,0);
       \draw[->,-latex,ultra thick] (\stepsize*2,\stepsize) --++ (\arrowsize,0);
       \draw[->,-latex,ultra thick] (\stepsize,\stepsize) --++ (0,\arrowsize);
       \draw[->,-latex,ultra thick] (0,\stepsize) --++ (0,\arrowsize);
       \draw[->,-latex,ultra thick] (\stepsize*2,\stepsize) --++ (0,\arrowsize);
       \node[align=center] at (\stepsize*0.5,\stepsize*0.5) {$D_\infty$};
       \node[align=center] at (\stepsize*1.5,\stepsize*0.5) {$D_\infty - \epsilon$};
       \draw[thin] (0,\stepsize + \arrowsize*1.5) --++ (\stepsize*2.0,0) node[align=center,pos=0.25,above]{$\kappa L$} node[align=center,pos=0.75,above]{$(1-\kappa) L$};
       \draw[thin] (0,\stepsize + \arrowsize*1.3) --++ (0,\arrowsize*0.4);
       \draw[thin] (\stepsize,\stepsize + \arrowsize*1.3) --++ (0,\arrowsize*0.4);
       \draw[thin] (\stepsize*2,\stepsize + \arrowsize*1.3) --++ (0,\arrowsize*0.4);
\end{tikzpicture}}
	\caption{Boundary value problem for the biaxial mesh sensitivity test with disturbance \(\varepsilon = 10^{-5}\) in the asymptotic damage limit parameter in the right element.
		The parameter \(\kappa\) characterizes length of the elements.}
	\label{fig:biaxial-bvp}
\end{figure}

\begin{figure}
	\centering
	\begin{subfigure}[b]{0.48\textwidth}
		\centering
		\ifthenelse{\boolean{professormode}}
		{\includegraphics{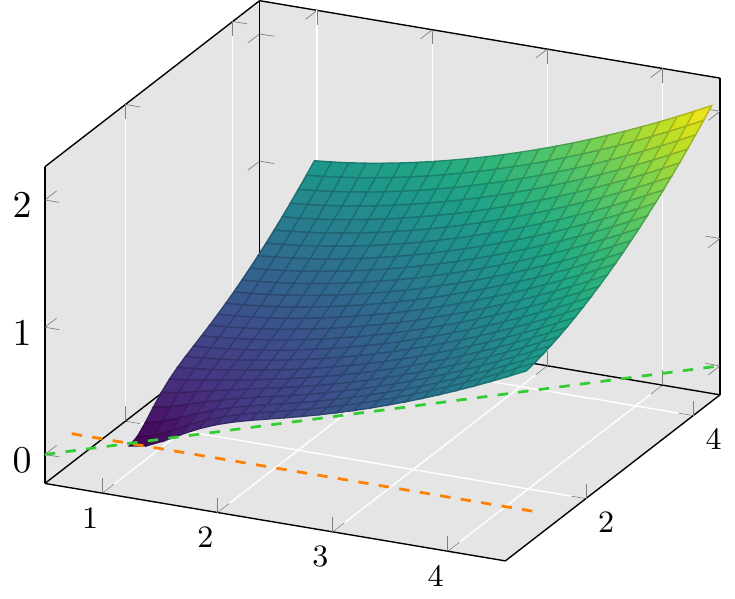}}
		{\input{figures/tikz/multid-biaxial-energy-nh-w.tex}}
		\caption{Non-convex \(W\) in \(F_{11}, F_{22}\) plane}
		\label{fig:multid-biaxial-energy-nh-w}
	\end{subfigure}
	\hfill
	\begin{subfigure}[b]{0.48\textwidth}
		\centering
		\ifthenelse{\boolean{professormode}}
		{\includegraphics{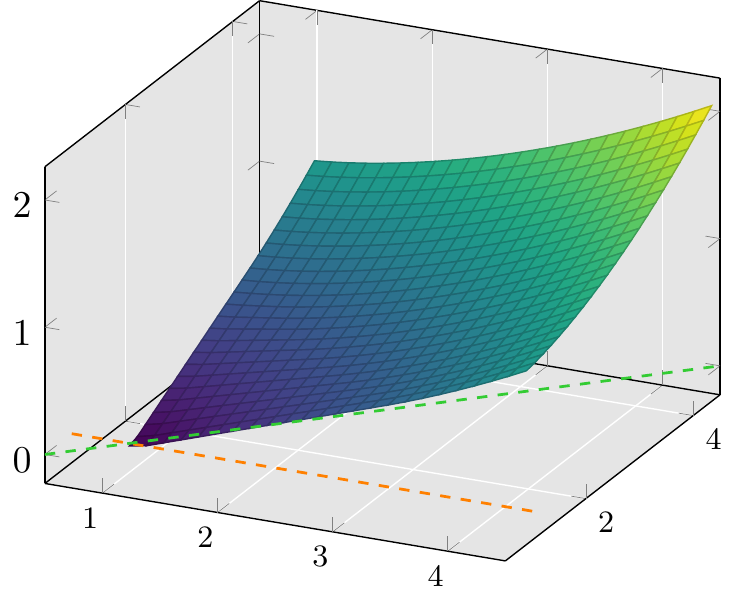}}
		{\input{figures/tikz/multid-biaxial-energy-nh-wrc.tex}}
		\caption{Approximation $W^{15}_{\delta,r}$ in \(F_{11}, F_{22}\) plane}
		\label{fig:multid-biaxial-energy-nh-wrc}
	\end{subfigure}
	\begin{subfigure}[b]{0.48\textwidth}
		\centering
		\ifthenelse{\boolean{professormode}}
		{\includegraphics{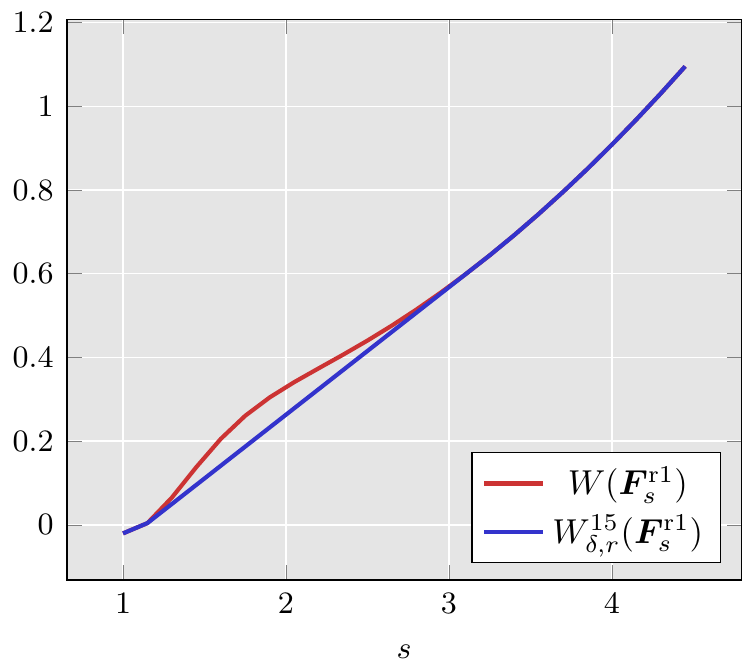}}
		{\tikzsetfigurename{multid-biaxial-energy-nh-rk1line}
\begin{tikzpicture}
	\begin{axis}[
		xlabel style={font={\footnotesize}}, 
		ylabel style={font={\footnotesize}}, 
		legend style={font={\small}}, 
		yticklabel style={font={\footnotesize}}, 
		xticklabel style={font={\footnotesize}}, 
		axis background/.style={fill={white!89.803921568!black}}, 
		x grid style={white}, 
		y grid style={white}, 
		xmajorgrids, 
		ymajorgrids, 
		legend pos={south east},
		colormap/viridis, 
		unbounded coords={jump}, 
		xlabel={$s$},
		]
		\addplot[color={red!60!gray}, very thick]
		coordinates {
					(1.0,-0.02042332301005384)
					(1.15,0.004255671569350972)
					(1.3,0.06531821608971167)
					(1.45,0.13810422558869992)
					(1.6,0.2055006823480171)
					(1.75,0.2607121165073966)
					(1.9,0.30450948069910494)
					(2.05,0.34087380715048615)
					(2.2,0.37389258620334603)
					(2.35,0.406480946078329)
					(2.5,0.4403074691309739)
					(2.65,0.47617294132323124)
					(2.8,0.5143998355635226)
					(2.95,0.55508960892883)
					(3.1,0.5982567261869194)
					(3.25,0.6438872075303925)
					(3.4,0.691960544355519)
					(3.55,0.7424566599868838)
					(3.7,0.7953576178157784)
					(3.85,0.8506477488961359)
					(4.0,0.9083133991249472)
					(4.15,0.9683426337620384)
					(4.3,1.0307249762617803)
					(4.45,1.0954511903135835)
				}
		;
		\addlegendentry {$W(\boldsymbol{F}^{\text{r1}}_{s})$}
		\addplot[color={blue!60!gray}, very thick]
		coordinates {
					(1.0,-0.02042332301005384)
					(1.15,0.00425567156935076)
					(1.3,0.049943638545469925)
					(1.45,0.0956316054989016)
					(1.6,0.1413195722213957)
					(1.75,0.18700753915471807)
					(1.9,0.23269550588185733)
					(2.05,0.27838347282200604)
					(2.2,0.32407143956989304)
					(2.35,0.3697594065007619)
					(2.5,0.4154473732693808)
					(2.65,0.4611353401909901)
					(2.8,0.5068233070412939)
					(2.95,0.5525112738926911)
					(3.1,0.5981992407228856)
					(3.25,0.6438872075303925)
					(3.4,0.691960544355519)
					(3.55,0.7424566599868838)
					(3.7,0.7953576178157786)
					(3.85,0.8506477488961359)
					(4.0,0.9083133991249472)
					(4.15,0.9683426337620387)
					(4.3,1.0307249762617803)
					(4.45,1.0954511903135835)
				}
		;
		\addlegendentry {$W^{15}_{\delta,r}(\boldsymbol{F}^{\text{r1}}_{s})$}
	\end{axis}
\end{tikzpicture}}
		\caption{{\color{orange} Rank-one line} \(\boldsymbol{F}^{\text{r1}}_s\)}
		\label{fig:multid-biaxial-energy-nh-rk1line}
	\end{subfigure}
	\hfill
	\begin{subfigure}[b]{0.48\textwidth}
		\centering
		\ifthenelse{\boolean{professormode}}
		{\includegraphics{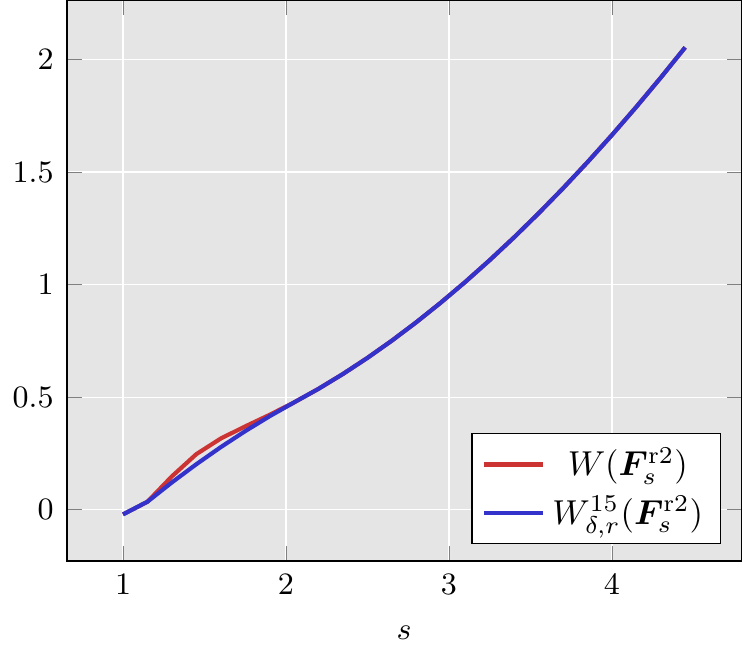}}
		{\tikzsetfigurename{multid-biaxial-energy-nh-rk2line}
\begin{tikzpicture}
	\begin{axis}[
		xlabel style={font={\footnotesize}}, 
		ylabel style={font={\footnotesize}}, 
		legend style={font={\small}}, 
		yticklabel style={font={\footnotesize}}, 
		xticklabel style={font={\footnotesize}}, 
		axis background/.style={fill={white!89.803921568!black}}, 
		x grid style={white}, 
		y grid style={white}, 
		xmajorgrids, 
		ymajorgrids, 
		legend pos={south east},
		colormap/viridis, 
		unbounded coords={jump}, 
		xlabel={$s$},
		]
		\addplot[color={red!60!gray}, very thick]
		coordinates {
						(1.0,-0.02042332301005384)
						(1.15,0.03593262552095711)
						(1.3,0.14858248018580514)
						(1.45,0.24720601687466437)
						(1.6,0.3166174047420501)
						(1.75,0.37044419773074827)
						(1.9,0.42190222677466777)
						(2.05,0.47697479925081315)
						(2.2,0.5373545632789718)
						(2.35,0.603265022161523)
						(2.5,0.6745980724205369)
						(2.65,0.7512122406720787)
						(2.8,0.8329852639634654)
						(2.95,0.9198167424592028)
						(3.1,1.0116241854098906)
						(3.25,1.1083392946935977)
						(3.4,1.2099050783016747)
						(3.55,1.3162736362226728)
						(3.7,1.4274044466892548)
						(3.85,1.5432630255786368)
						(4.0,1.6638198672652138)
						(4.15,1.789049600172466)
						(4.3,1.9189303078264814)
						(4.45,2.0534429787248327)
					}
		;
		\addlegendentry {$W(\boldsymbol{F}^{\text{r2}}_{s})$}
		\addplot[color={blue!60!gray}, very thick]
		coordinates {
						(1.0,-0.02042332301005384)
						(1.15,0.03593262552095686)
						(1.3,0.12160542193541421)
						(1.45,0.20239237927412676)
						(1.6,0.2783020323175994)
						(1.75,0.34937325119806073)
						(1.9,0.41559837343257006)
						(2.05,0.4769747992508126)
						(2.2,0.5373545632789718)
						(2.35,0.603265022161523)
						(2.5,0.6745980724205369)
						(2.65,0.7512122406720787)
						(2.8,0.8329852639634654)
						(2.95,0.9198167424592028)
						(3.1,1.0116241854098906)
						(3.25,1.1083392946935977)
						(3.4,1.2099050783016747)
						(3.55,1.3162736362226728)
						(3.7,1.4274044466892553)
						(3.85,1.5432630255786368)
						(4.0,1.6638198672652138)
						(4.15,1.7890496001724665)
						(4.3,1.9189303078264814)
						(4.45,2.0534429787248327)
					}
		;
		\addlegendentry {$W^{15}_{\delta,r}(\boldsymbol{F}^{\text{r2}}_{s})$}
	\end{axis}
\end{tikzpicture}}
		\caption{{\color{green!60!gray}Rank-two line} \(\boldsymbol{F}^{\text{r2}}_s\)}
		\label{fig:multid-biaxial-energy-nh-rk2line}
	\end{subfigure}
	\caption{Non-convex incremental stress potential (\subref{fig:multid-biaxial-energy-nh-w}) and the associated rank-one convex envelope approximation $W^k_{\delta,r}$ (\subref{fig:multid-biaxial-energy-nh-wrc}) for a Neo-Hooke effective strain energy density \(\psi^{0}\).
	The value \(s\) corresponds to the line characterizing equations \eqref{eq:rankOneRankTwoLine}, where (\subref{fig:multid-biaxial-energy-nh-rk1line}) shows the rank-one path and (\subref{fig:multid-biaxial-energy-nh-rk2line}) the rank-two path, respectively.
	Note that the rank-one convex envelope is convex along the rank-one line but non-convex following the rank-two line.}
	\label{fig:multid-biaxial-energy-nh}
\end{figure}

\begin{figure}
	\centering
	\begin{subfigure}[b]{0.48\textwidth}
		\centering
		\ifthenelse{\boolean{professormode}}
		{\includegraphics{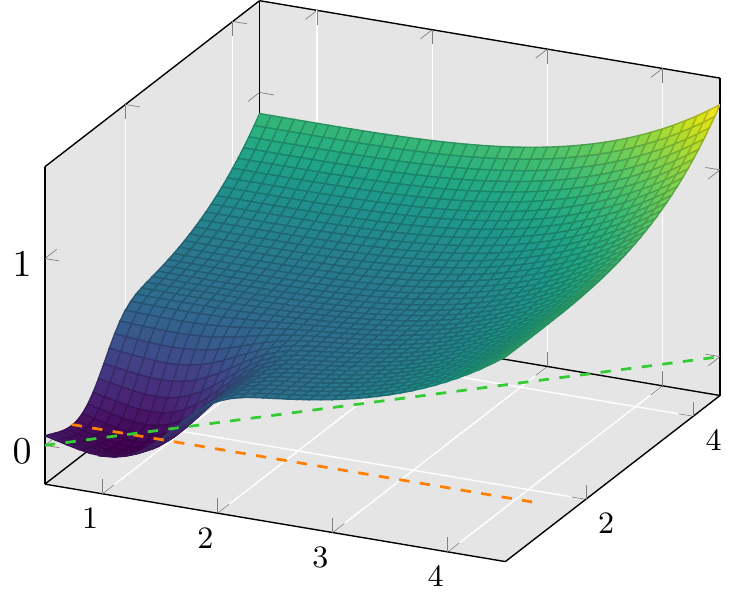}}
		{\input{figures/tikz/multid-biaxial-energy-svk-w.tex}}
		\caption{Non-convex \(W\) in \(F_{11}, F_{22}\) plane}
		\label{fig:multid-biaxial-energy-svk-w}
	\end{subfigure}
	\hfill
	\begin{subfigure}[b]{0.48\textwidth}
		\centering
		\ifthenelse{\boolean{professormode}}
		{\includegraphics{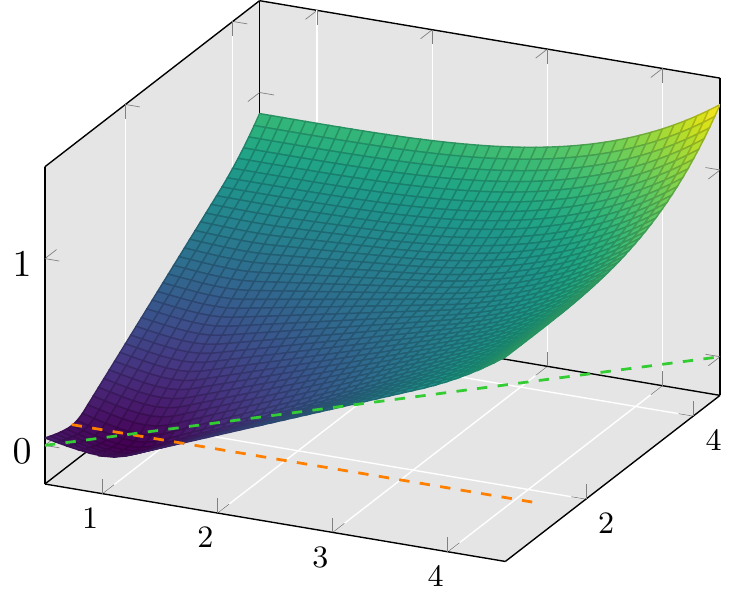}}
		{\input{figures/tikz/multid-biaxial-energy-svk-wrc.tex}}
		\caption{Approximation $W^{11}_{\delta,r}$ in \(F_{11}, F_{22}\) plane}
		\label{fig:multid-biaxial-energy-svk-wrc}
	\end{subfigure}
	\begin{subfigure}[b]{0.48\textwidth}
		\centering
		\ifthenelse{\boolean{professormode}}
		{\includegraphics{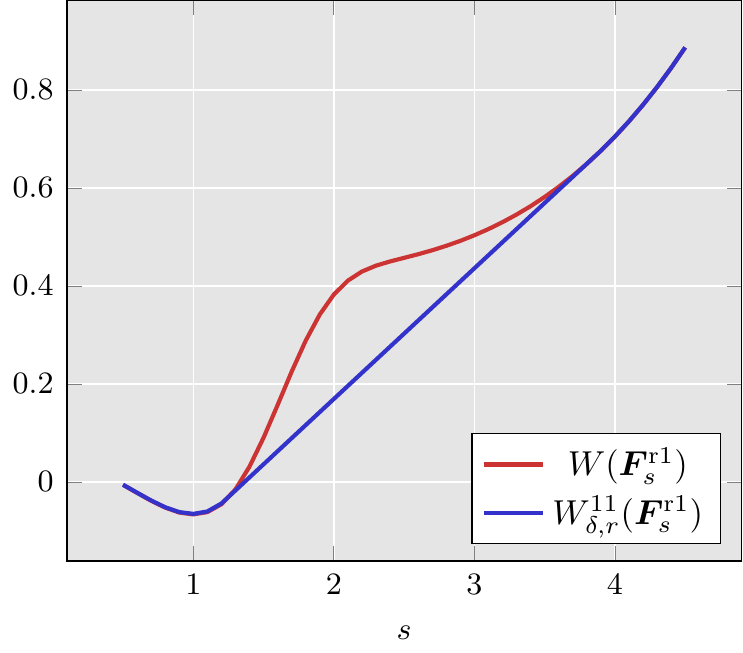}}
		{\tikzsetfigurename{multid-biaxial-energy-svk-rk1line}
\begin{tikzpicture}
	\begin{axis}[
		xlabel style={font={\footnotesize}},
		ylabel style={font={\footnotesize}}, 
		legend style={font={\small}}, 
		yticklabel style={font={\footnotesize}}, 
		xticklabel style={font={\footnotesize}}, 
		axis background/.style={fill={white!89.803921568!black}}, 
		x grid style={white}, 
		y grid style={white}, 
		xmajorgrids, 
		ymajorgrids, 
		legend pos={south east},
		colormap/viridis, 
		unbounded coords={jump}, 
		xlabel={$s$}
		]
		\addplot[color={red!60!gray}, very thick]
		coordinates {
				(0.5,-0.006043362538320052)
				(0.6,-0.02244373203736183)
				(0.7,-0.03847941052072823)
				(0.8,-0.052477109798850896)
				(0.9,-0.06250611272991957)
				(1.0,-0.06637827321988299)
				(1.1,-0.06164801622244842)
				(1.2,-0.04561233773908191)
				(1.3,-0.01531080481900804)
				(1.4,0.03128985415120766)
				(1.5,0.09029407234634235)
				(1.6,0.15692734266542063)
				(1.7,0.22507920513404356)
				(1.8,0.28834946339303213)
				(1.9,0.3416346142110589)
				(2.0,0.38235208503827534)
				(2.1,0.41076399814032283)
				(2.2,0.42930097988143157)
				(2.3,0.4413092649400827)
				(2.4,0.4498804235912073)
				(2.5,0.457224273663809)
				(2.6,0.46462299454584244)
				(2.7,0.47270476581818865)
				(2.8,0.481752580949693)
				(2.9,0.4919092853997542)
				(3.0,0.5032746454914863)
				(3.1,0.5159398217724034)
				(3.2,0.5299967009304192)
				(3.3,0.5455398261845188)
				(3.4,0.5626667011960107)
				(3.5,0.58147782619639)
				(3.6,0.6020767011963999)
				(3.7,0.6245698261964003)
				(3.8,0.6490667011963973)
				(3.9,0.6756798261963981)
				(4.0,0.7045247011963989)
				(4.1,0.7357198261963989)
				(4.2,0.7693867011963968)
				(4.3,0.8056498261964018)
				(4.4,0.8446367011964009)
				(4.5,0.8864778261963986)
			}
		;
		\addlegendentry {$W(\boldsymbol{F}^{\text{r1}}_{s})$}
		\addplot[color={blue!60!gray}, very thick]
		coordinates {
				(0.5,-0.0057738666039424125)
				(0.6,-0.022056247833206807)
				(0.7,-0.03795248563410924)
				(0.8,-0.05178929181708103)
				(0.9,-0.06163594924031218)
				(1.0,-0.0653043118097511)
				(1.1,-0.06034880447910561)
				(1.2,-0.0440664232498412)
				(1.3,-0.017402219347409296)
				(1.4,0.009261984555022524)
				(1.5,0.035926188457454517)
				(1.6,0.06259039235988646)
				(1.7,0.0892545962623183)
				(1.8,0.11591880016475026)
				(1.9,0.1425830040671821)
				(2.0,0.16924720796961412)
				(2.1,0.19591141187204605)
				(2.2,0.222575615774478)
				(2.3,0.24923981967690972)
				(2.4,0.2759040235793417)
				(2.5,0.30256822748177375)
				(2.6,0.3292324313842057)
				(2.7,0.35589663528663773)
				(2.8,0.3825608391890695)
				(2.9,0.4092250430915014)
				(3.0,0.4358892469939335)
				(3.1,0.4625534508963654)
				(3.2,0.4892176547987974)
				(3.3,0.5158818587012292)
				(3.4,0.5425460626036611)
				(3.5,0.5692102665060932)
				(3.6,0.5958744704085251)
				(3.7,0.6225386743109571)
				(3.8,0.6492028782133888)
				(3.9,0.6758699668213977)
				(4.0,0.7047247168213984)
				(4.1,0.7359299668213961)
				(4.2,0.7696072168213997)
				(4.3,0.8058809668213968)
				(4.4,0.8448787168213991)
				(4.5,0.8867309668213972)
			}
		;
		\addlegendentry {$W^{11}_{\delta,r}(\boldsymbol{F}^{\text{r1}}_{s})$}
	\end{axis}
\end{tikzpicture}}
		\caption{{\color{orange} Rank-one line} \(\boldsymbol{F}^{\text{r1}}_s\)}
		\label{fig:multid-biaxial-energy-svk-rk1line}
	\end{subfigure}
	\hfill
	\begin{subfigure}[b]{0.48\textwidth}
		\centering
		\ifthenelse{\boolean{professormode}}
		{\includegraphics{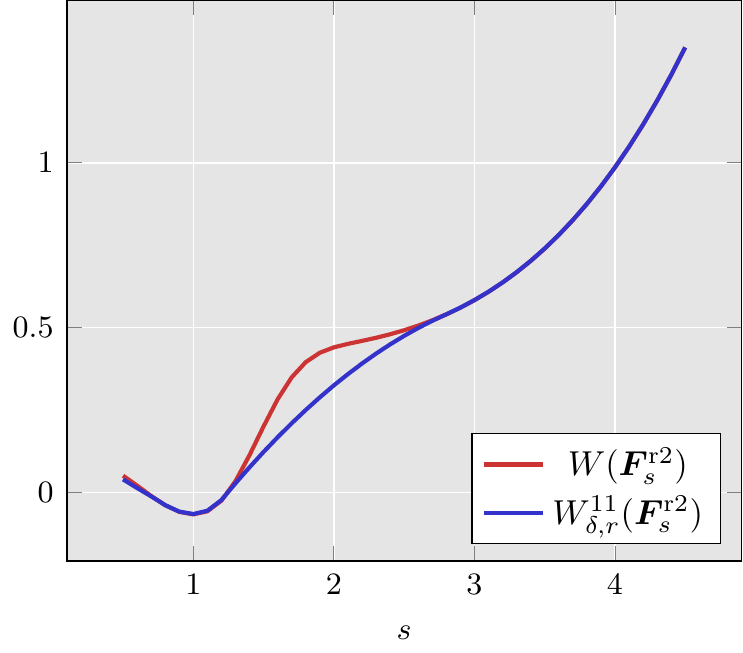}}
		{\tikzsetfigurename{multid-biaxial-energy-svk-rk2line}
\begin{tikzpicture}
	\begin{axis}[
		xlabel style={font={\footnotesize}},
		ylabel style={font={\footnotesize}}, 
		legend style={font={\small}}, 
		yticklabel style={font={\footnotesize}}, 
		xticklabel style={font={\footnotesize}}, 
		axis background/.style={fill={white!89.803921568!black}}, 
		x grid style={white}, 
		y grid style={white}, 
		xmajorgrids, 
		ymajorgrids, 
		legend pos={south east},
		colormap/viridis, 
		unbounded coords={jump}, 
		xlabel={$s$}
		]
		\addplot[color={red!60!gray}, very thick]
		coordinates {
					(0.5,0.051035348211839526)
					(0.6,0.02096762142287706)
					(0.7,-0.010580547821573583)
					(0.8,-0.038575946377818915)
					(0.9,-0.058633952239956266)
					(1.0,-0.06637827321988299)
					(1.1,-0.056917759225013964)
					(1.2,-0.024846402258280942)
					(1.3,0.034324126633894614)
					(1.4,0.11334318502542129)
					(1.5,0.20055343711944906)
					(1.6,0.2827513749545163)
					(1.7,0.3492325727532076)
					(1.8,0.39554262046245836)
					(1.9,0.4238848254426175)
					(2.0,0.4402757479099684)
					(2.1,0.45086722025652204)
					(2.2,0.4598277882239854)
					(2.3,0.46923503911720044)
					(2.4,0.47991275047151394)
					(2.5,0.49218043894356445)
					(2.6,0.5062186702805702)
					(2.7,0.5221849499276674)
					(2.8,0.5402387011621022)
					(2.9,0.5605449511958076)
					(3.0,0.5832747011963919)
					(3.1,0.6086049511963987)
					(3.2,0.636718701196398)
					(3.3,0.6678049511963993)
					(3.4,0.702058701196397)
					(3.5,0.7396809511963989)
					(3.6,0.7808787011964009)
					(3.7,0.8258649511964018)
					(3.8,0.8748587011963957)
					(3.9,0.9280849511963973)
					(4.0,0.9857747011963989)
					(4.1,1.0481649511963989)
					(4.2,1.1154987011963948)
					(4.3,1.1880249511964047)
					(4.4,1.265998701196403)
					(4.5,1.3496809511963983)
				}
		;
		\addlegendentry {$W(\boldsymbol{F}^{\text{r2}}_{s})$}
		\addplot[color={blue!60!gray}, very thick]
		coordinates {
					(0.5,0.03973331853132353)
					(0.6,0.013586631182548421)
					(0.7,-0.011610827419134688)
					(0.8,-0.03827427182441101)
					(0.9,-0.05796758667087315)
					(1.0,-0.0653043118097511)
					(1.1,-0.055393297148459725)
					(1.2,-0.022828534689931273)
					(1.3,0.027856513426636136)
					(1.4,0.07647731138997994)
					(1.5,0.12303385920010067)
					(1.6,0.1675261568569978)
					(1.7,0.20995420436067125)
					(1.8,0.2503180017111216)
					(1.9,0.2886175489083481)
					(2.0,0.3248528459523514)
					(2.1,0.3590238928431312)
					(2.2,0.3911306895806874)
					(2.3,0.42117323616501995)
					(2.4,0.44915153259612933)
					(2.5,0.47506557887401535)
					(2.6,0.49891537499867766)
					(2.7,0.5207009209701168)
					(2.8,0.5404222167883326)
					(2.9,0.5607427168208314)
					(3.0,0.5834872168213918)
					(3.1,0.6088327168213992)
					(3.2,0.6369622168213995)
					(3.3,0.6680647168213989)
					(3.4,0.7023352168213988)
					(3.5,0.7399747168213979)
					(3.6,0.7811902168213978)
					(3.7,0.8261947168213988)
					(3.8,0.8752072168213956)
					(3.9,0.9284527168214014)
					(4.0,0.9861622168213993)
					(4.1,1.0485727168213983)
					(4.2,1.1159272168213983)
					(4.3,1.188474716821394)
					(4.4,1.2664702168213973)
					(4.5,1.3501747168213933)
				}
		;
		\addlegendentry {$W^{11}_{\delta,r}(\boldsymbol{F}^{\text{r2}}_{s})$}
	\end{axis}
\end{tikzpicture}}
		\caption{{\color{green!60!gray}Rank-two line} \(\boldsymbol{F}^{\text{r2}}_s\)}
		\label{fig:multid-biaxial-energy-svk-rk2line}
	\end{subfigure}
	\label{fig:multid-biaxial-energy-svk}
	\caption{The St.~Venant--Kirchhoff effective strain energy density and its associated incremental stress potential (\subref{fig:multid-biaxial-energy-svk-w}) and obtained rank-one convex envelope approximation (\subref{fig:multid-biaxial-energy-svk-wrc}).
	Again, the value \(s\) corresponds to the line characterizing equation \eqref{eq:rankOneRankTwoLine} and a similar behavior compared to Figure \ref{fig:multid-biaxial-energy-nh} can be observed.
	Namely, a convex behavior along the rank-one path (depicted in (\subref{fig:multid-biaxial-energy-svk-rk1line}))) while a lower and yet noticable non-convexity is left along the rank-two path, shown in (\subref{fig:multid-biaxial-energy-svk-rk2line}).}
\end{figure}

\begin{figure}
\begin{subfigure}[b]{0.49\textwidth} \label{fig:multid-biaxial-neohooke}
	 \ifthenelse{\boolean{professormode}}
	 {\includegraphics{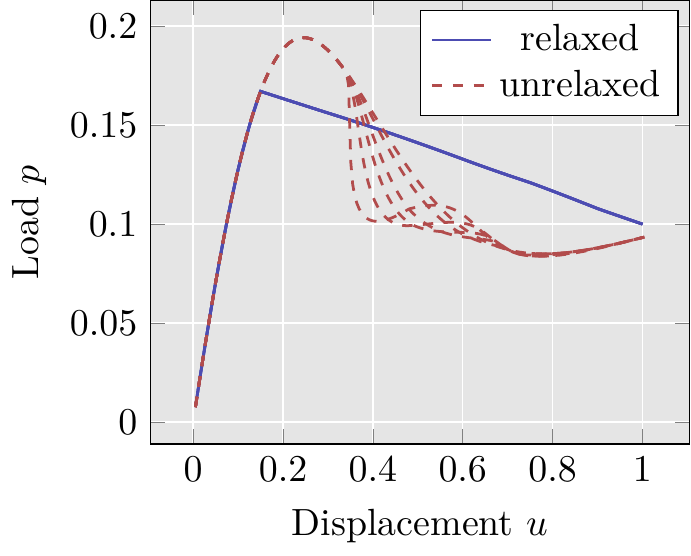}}
	 {\input{figures/tikz/multid-biaxial-neohooke}}
	 \label{fig:biaxial-nh}
	 \caption{}
\end{subfigure}
\hfill
\begin{subfigure}[b]{0.49\textwidth} \label{fig:multid-biaxial-svk}
	 \ifthenelse{\boolean{professormode}}
	 {\includegraphics{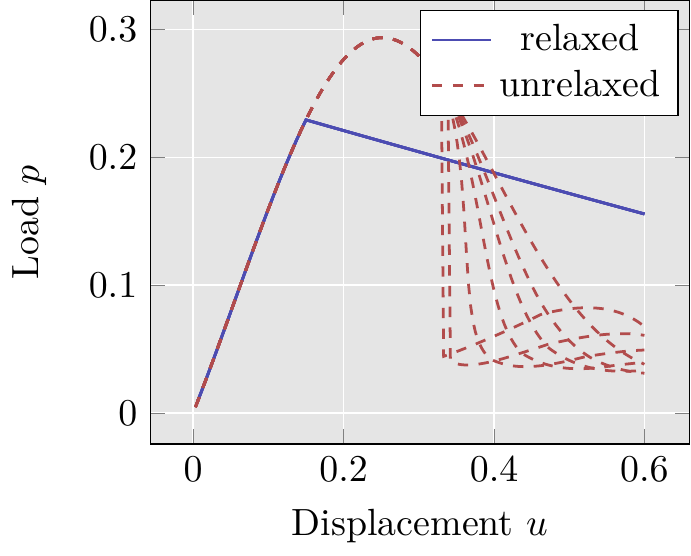}}
	 {\input{figures/tikz/multid-biaxial-svk}}
	 \label{fig:biaxial-svk}
	 \caption{}
\end{subfigure}
\caption{Biaxial force--displacement response for Neo-Hooke effective strain energy density at the left-hand side and St.~Venant--Kirchhoff effective strain energy density at the right-hand side, respectively.
The relaxed model shows, despite the strain softening behavior, a mesh independent response; in contrast, the unrelaxed model shows the typical mesh dependence.}
\label{fig:biaxial-response}
\end{figure}

The biaxial tension boundary value problem illustrated in Figure \ref{fig:biaxial-bvp} is used for a further mesh sensitivity tests.
The damage limit parameter in the right element is again disturbed by a small parameter \(\epsilon = 10^{-5}\) and the discretization of the individual elements is varied by the parameter \({\kappa \in \{0.3, 0.4, 0.5, 0.6, 0.7, 0.8, 1.0\}}\). 
The disturbance parameter \(\epsilon\) is applied for varying \(\kappa\) to the right element.
Material and convexification parameters can be found in Table \ref{tab:biaxialperturbation}.
\begin{table}
    \smaller
	\begin{tabular}{|c|c|c|c|c|c|c|c|c|c|c|c|}
		\hline
		$\psi^0$ & $\lambda$ & $\mu$ & $D_0$ & $D_\infty$ & $\delta_{ij}$ & $F_{ii}^{\text{min}}$ & $F_{ii}^{\text{max}}$ & $F_{ij}^{\text{min}} \, \forall i\neq j$  & $F_{ij}^{\text{max}} \, \forall i\neq j$ & $|\mathcal{N}|$\\
		\hline
		NH & 0.5 & 1.0 & 0.3 & 0.9 & 0.15 & 1.0 & 3.4 & -0.15 & 0.15 & 2601\\
		\hline
		STVK & 0.5 & 1.0 & 0.4 & 0.99 & 0.15 & 1.0 & 3.1 & -0.15 & 0.15 & 2025\\
		\hline
	\end{tabular}
    \caption{Parameter settings for the biaxial perturbation test.
             Here, the convexification grid is specially constructed by coarsely discretizing the off-diagonal components of the deformation gradient.}
    \label{tab:biaxialperturbation}
\end{table}
In contrast to the material point study, the relaxation is performed in each non-convex step until the convexified regime is reached.
After that, the approximated rank-one convex envelope is fixed and reused in subsequent incremental time steps.
The Figures \ref{fig:multid-biaxial-energy-nh-w} and \ref{fig:multid-biaxial-energy-svk-w} show the unrelaxed and the Figures \ref{fig:multid-biaxial-energy-nh-wrc} and \ref{fig:multid-biaxial-energy-svk-wrc} the relaxed (and then fixed) incremental stress potential along rank-one and rank-two lines which are given by the paths
\begin{align} \label{eq:rankOneRankTwoLine}
	\boldsymbol{F}_s^{\text{r1}} = \begin{bmatrix} 1 & 0 \\ 0 & 1 \end{bmatrix} + s \begin{bmatrix} 1 & 0 \\ 0 & 0 \end{bmatrix} \quad \text{ and } \quad \boldsymbol{F}_s^{\text{r2}} = \begin{bmatrix} 1 & 0 \\ 0 & 1 \end{bmatrix} + s \begin{bmatrix} 1 & 0 \\ 0 & 1 \end{bmatrix}
\end{align}
for varying \(s\). 
Along the rank-two line (which corresponds to the evolution of the deformation of this boundary value problem), non-convexity of the deformation path is visible. 
Indeed, the non-convexity along the paths is required to reflect some strain softening (since a decreasing slope is required) as illustrated in Figure \ref{fig:biaxial-response}.

This illustration shows the resulting material responses of the biaxial test using STVK and NH effective strain energy densities.
In both cases, the derivative was computed with the novel tree decomposition procedure.
It was not possible to obtain a convergent finite element solver by using the subdifferential description of the derivative.
Since the tree decomposition approach is also capable of describing tangent moduli (which may vanish across rank-one lines), a Newton scheme with Armijo--Goldstein based line search was applied for the St.~Venant--Kirchhoff case.
As pointed out in \cite[Remark 9.6]{Bar:2015:nmn}, the second derivative can vanish for semi-convex envelopes; however, in this special case rank-two deformation paths are expected and, thus, the second derivative does not vanish.
In the Neo-Hooke effective strain energy density case, the descent method with line search as in the uniaxial example was used again.
	
The unrelaxed model shows strain softening, but also considerably strong mesh sensitivity.
Despite the observed non-convexity along the rank-two line, the response is mesh independent.
Note that although some strain softening is observed along specific lines, the relaxation approach presented here with fixed convex envelopes does not represent a suitable model for the description of a realistic strain softening.
The reason is two-fold: (i) strain-softening cannot be modeled specifically, and (ii) strain-softening cannot be obtained along rank-one lines, which, however, may be particularly relevant in practice.
If specifically strain-softening needs to be modeled, the concept of the approach in \cite{KohBal:2023:emr} can be considered, where the convex hull is recomputed in each time step.

\subsection{Lamination depth convergence study} \label{sec:examples:laminationdepth}
\begin{figure} 
	\begin{subfigure}[b]{0.49\textwidth}
		\centering
		\ifthenelse{\boolean{professormode}}
		{\includegraphics{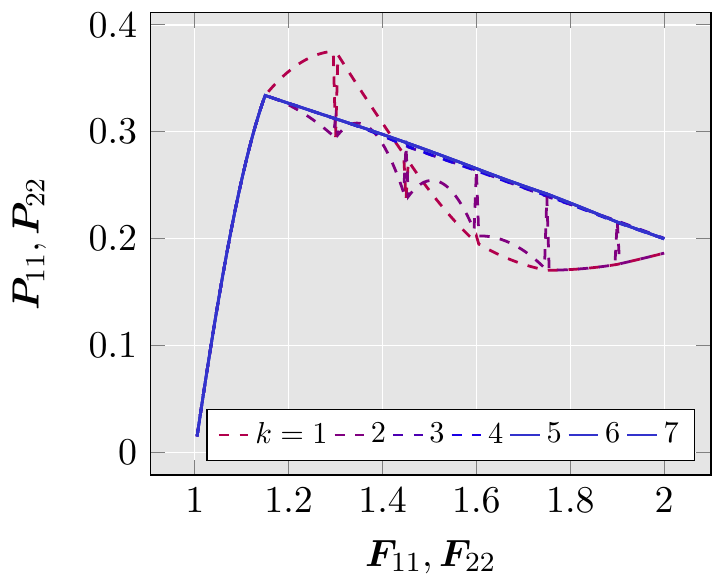}}
		{\input{figures/tikz/treedepth-neohooke-biaxial.tex}}
	\end{subfigure}
	\hfill
	\begin{subfigure}[b]{0.49\textwidth}
		\centering
		\ifthenelse{\boolean{professormode}}
		{\includegraphics{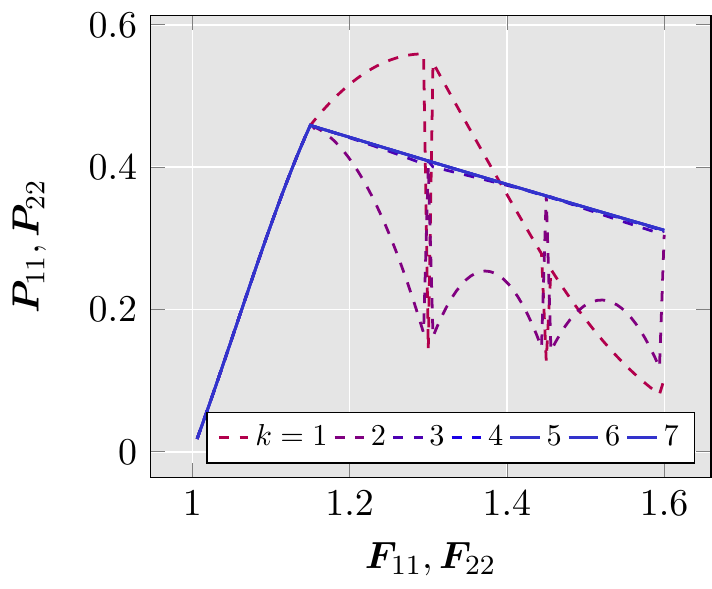}}
		{\input{figures/tikz/treedepth-svk-biaxial.tex}}
	\end{subfigure}
	\caption{Lamination depth study in terms of the obtained first Piola--Kirchhoff stresses obtained by tree decomposition and evaluation.
	Here, the same material parameters as well as loading scenarios are evaluated as in the biaxial perturbation test.
	The left-hand side shows the Neo-Hooke and the right-hand side the St.~Venant--Kirchhoff effective strain energy density, respectively.
	The stresses converge to a smooth function for a lamination depth $k\geq 5$ that includes strain softening.}
	\label{fig:treedepth}
\end{figure}

As already mentioned, the number of global iterations of the convexification algorithm can be interpreted as the number of successive rank-one laminations, and hence the number of laminates that can occur in the microstructure.
In Figure \ref{fig:treedepth}, the response of the biaxial boundary value problem of the previous section is shown in relation to the lamination depth (or number of global lamination iterations in the convexification algorithm).
While so far, only criteria with respect to the function values were discussed, we discuss in this section the convergence in terms of the first Piola--Kirchhoff stresses $\boldsymbol{P}$.
The stresses serve as a better convergence indicator, since their value will be used in the finite element solver.
Only in special cases, e.g. line searches, the rank-one convex envelope function value is needed.
Since the hull is fixed, the evolution path of the deformation gradient in the biaxial setting somehow crosses some left non-convexities which could be an explanation for the buckling. 
Interestingly, at some points, the approximation coincides with the converged response.
It is recognizable that 5 global iterations already deliver a suitable convex hull since the higher lamination depths do not produce a notable change in material response.
In other words, the convexification procedure delivered a meaningful material response after only five iterations.

\subsection{Triaxial test} \label{sec:examples:triaxial}
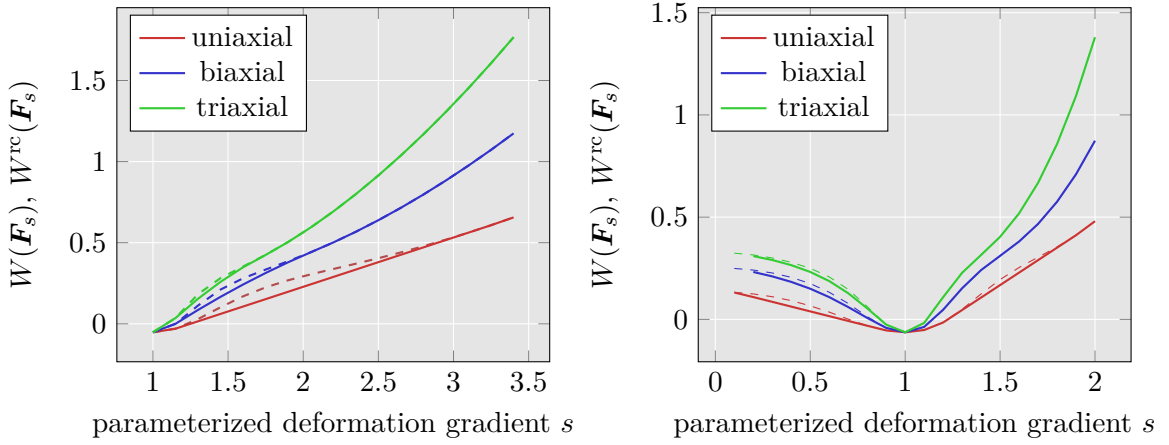
\begin{figure} 
    \begin{subfigure}[b]{0.49\textwidth}
    \centering
	\ifthenelse{\boolean{professormode}}
	{\includegraphics{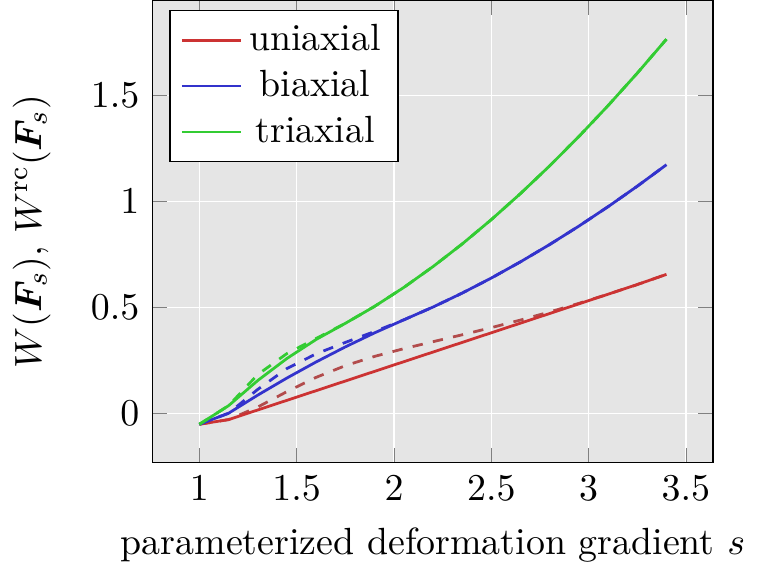}}
    {\tikzsetfigurename{3dconvexification-lines}
\begin{tikzpicture}
\begin{axis}[axis background/.style={{fill={white!89.803921568!black}}},
width=0.99\textwidth,
x grid style={{white}},
y grid style={{white}},
xmajorgrids,
ymajorgrids,
ylabel={{$W(\boldsymbol{F}_s)$, $W^{\text{rc}}(\boldsymbol{F}_s)$}},
xlabel={{parameterized deformation gradient $s$}},
legend style={{at={(0.03,0.65)},
anchor=south west}}]
    \addplot+[no marks, dashed, thick, red!40!gray, forget plot]
        table[row sep={\\}]
        {
            \\
            1.0  -0.0519271582295569  \\
            1.15  -0.03001984675045921  \\
            1.3  0.029386469143897476  \\
            1.45  0.10217247864288567  \\
            1.6  0.16956893540220302  \\
            1.75  0.22478036956158248  \\
            1.9  0.26857773375329075  \\
            2.05  0.30494206020467207  \\
            2.2  0.33796083925753184  \\
            2.35  0.37054919913251483  \\
            2.5  0.4043757221851597  \\
            2.65  0.4402411943774166  \\
            2.8  0.47846808861770806  \\
            2.95  0.5191578619830154  \\
            3.1  0.5623249792411048  \\
            3.25  0.607955460584578  \\
            3.4  0.6560287974097049  \\
        }
        ;
    \addplot+[no marks, solid, thick, red!60!gray]
        table[row sep={\\}]
        {
            \\
            1.0  -0.0519271582295569  \\
            1.15  -0.030019846750459347  \\
            1.3  0.015545139865247207  \\
            1.45  0.061110126480844795  \\
            1.6  0.10667511309440862  \\
            1.75  0.15224009970974356  \\
            1.9  0.1978050863236695  \\
            2.05  0.24337007293882895  \\
            2.2  0.2889350595530263  \\
            2.35  0.3345000461680092  \\
            2.5  0.3800650327824802  \\
            2.65  0.42563001939728873  \\
            2.8  0.47119500601238895  \\
            2.95  0.516759992626668  \\
            3.1  0.5623249792411049  \\
            3.25  0.607955460584578  \\
            3.4  0.6560287974097049  \\
        }
        ;
    \addlegendentry {uniaxial}
    \addplot+[no marks, dashed, thick, blue!60!gray, forget plot]
        table[row sep={\\}]
        {
            \\
            1.0  -0.0519271582295569  \\
            1.15  8.785751429152278e-7  \\
            1.3  0.11265073323999095  \\
            1.45  0.2112742699288503  \\
            1.6  0.280685657796236  \\
            1.75  0.33451245078493386  \\
            1.9  0.3859704798288536  \\
            2.05  0.44104305230499896  \\
            2.2  0.5014228163331572  \\
            2.35  0.5673332752157084  \\
            2.5  0.6386663254747237  \\
            2.65  0.7152804937262646  \\
            2.8  0.7970535170176513  \\
            2.95  0.8838849955133886  \\
            3.1  0.9756924384640774  \\
            3.25  1.0724075477477846  \\
            3.4  1.1739733313558616  \\
        }
        ;
    \addplot+[no marks, solid, thick, blue!60!gray]
        table[row sep={\\}]
        {
            \\
            1.0  -0.0519271582295569  \\
            1.15  8.785751426873361e-7  \\
            1.3  0.0856736296323862  \\
            1.45  0.166453445731858  \\
            1.6  0.2423683659400153  \\
            1.75  0.3134297663825032  \\
            1.9  0.37965416898428234  \\
            2.05  0.44104305230499896  \\
            2.2  0.5014228163331574  \\
            2.35  0.5673332752157086  \\
            2.5  0.6386663254747237  \\
            2.65  0.7152804937262646  \\
            2.8  0.7970535170176513  \\
            2.95  0.8838849955133888  \\
            3.1  0.9756924384640774  \\
            3.25  1.0724075477477846  \\
            3.4  1.1739733313558616  \\
        }
        ;
    \addlegendentry {biaxial}
    \addplot+[no marks, dashed, thick, green!60!gray, forget plot]
        table[row sep={\\}]
        {
            \\
            1.0  -0.0519271582295569  \\
            1.15  0.03569458947117207  \\
            1.3  0.18273103458473483  \\
            1.45  0.28203840369377964  \\
            1.6  0.35334769786291276  \\
            1.75  0.42468857546675987  \\
            1.9  0.5045881155583215  \\
            2.05  0.59392916718535  \\
            2.2  0.6923031748541145  \\
            2.35  0.7992719370731448  \\
            2.5  0.9144860260916369  \\
            2.65  1.037670355656838  \\
            2.8  1.1686060942749523  \\
            2.95  1.3071172818555197  \\
            3.1  1.4530610695532356  \\
            3.25  1.6063204981638584  \\
            3.4  1.7667990765738888  \\
        }
        ;
    \addplot+[no marks, solid, thick, green!60!gray]
        table[row sep={\\}]
        {
            \\
            1.0  -0.0519271582295569  \\
            1.15  0.035694589471171725  \\
            1.3  0.15458382140858162  \\
            1.45  0.2588565941255296  \\
            1.6  0.34850481465492755  \\
            1.75  0.42468857546675987  \\
            1.9  0.5045881155583215  \\
            2.05  0.59392916718535  \\
            2.2  0.6923031748541149  \\
            2.35  0.7992719370731448  \\
            2.5  0.9144860260916369  \\
            2.65  1.037670355656838  \\
            2.8  1.1686060942749523  \\
            2.95  1.3071172818555203  \\
            3.1  1.4530610695532356  \\
            3.25  1.6063204981638584  \\
            3.4  1.7667990765738888  \\
        }
        ;
    \addlegendentry {triaxial}
\end{axis}
\end{tikzpicture}}
    \end{subfigure}
	\hfill
    \begin{subfigure}[b]{0.49\textwidth}
   	\centering
	\ifthenelse{\boolean{professormode}}
	{\includegraphics{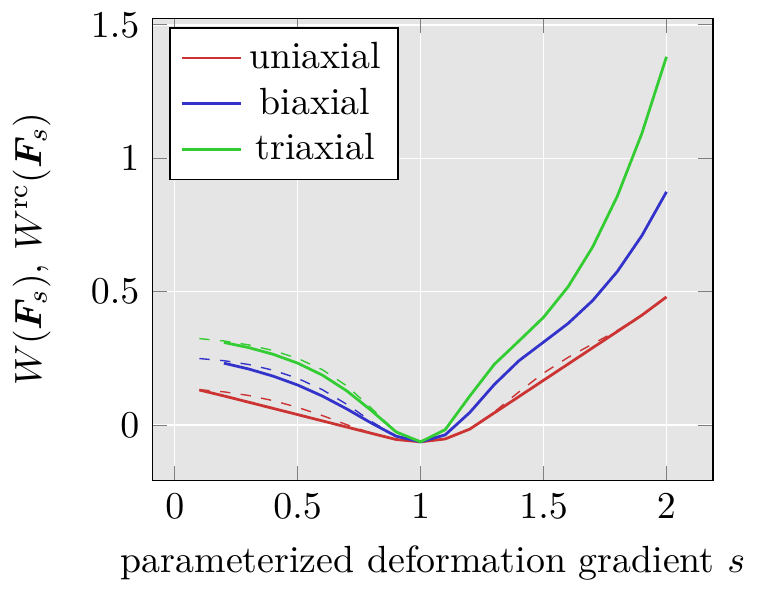}}
    {\tikzsetfigurename{3D-convexification-svk}
\begin{tikzpicture}
\begin{axis}[axis background/.style={{fill={white!89.803921568!black}}},
width=0.99\textwidth,
x grid style={{white}},
y grid style={{white}},
xmajorgrids,
ymajorgrids,
ylabel={{$W(\boldsymbol{F}_s)$, $W^{\text{rc}}(\boldsymbol{F}_s)$}},
xlabel={{parameterized deformation gradient $s$}},
legend style={{at={(0.03,0.65)},
anchor=south west}}]
    \addplot[mark={none}, thick, red!60!gray]
        table[row sep={\\}]
        {
            \\
            0.1  0.13132009674344483  \\
            0.2  0.1090643911902926  \\
            0.3  0.08570656531445904  \\
            0.4  0.0623487394386255  \\
            0.5  0.038990913562791954  \\
            0.6  0.01563308768695844  \\
            0.7  -0.007724738188875046  \\
            0.8  -0.03108256406470861  \\
            0.9  -0.05415520351355607  \\
            1.0  -0.06306346323658707  \\
            1.1  -0.052181073935045015  \\
            1.2  -0.015289527436620343  \\
            1.3  0.045689512576715885  \\
            1.4  0.10666855259005197  \\
            1.5  0.1676475926033878  \\
            1.6  0.22862663261672372  \\
            1.7  0.28960567263005893  \\
            1.8  0.35058471264339425  \\
            1.9  0.41156375265672884  \\
            2.0  0.48013641194030515  \\
        }
        ;
    \addlegendentry {uniaxial}
    \addplot[mark={none}, dashed, forget plot, red!60!gray]
        table[row sep={\\}]
        {
            \\
            0.1  0.13226803683925492  \\
            0.2  0.12432823365054202  \\
            0.3  0.11083082439678271  \\
            0.4  0.09149283894060084  \\
            0.5  0.06620980651454234  \\
            0.6  0.03548542225454793  \\
            0.7  0.0011181142789085108  \\
            0.8  -0.031082564064708607  \\
            0.9  -0.05415520351355607  \\
            1.0  -0.06306346323658707  \\
            1.1  -0.05218107393504501  \\
            1.2  -0.01528952743662021  \\
            1.3  0.04935751822530837  \\
            1.4  0.12432823365054202  \\
            1.5  0.19470955895868436  \\
            1.6  0.25355864866234484  \\
            1.7  0.3040003978940777  \\
            1.8  0.3542590348201135  \\
            1.9  0.41156375265672923  \\
            2.0  0.4801364119403057  \\
        }
        ;
    \addplot[mark={none}, thick, blue!60!gray]
        table[row sep={\\}]
        {
            \\
            0.1  nan  \\
            0.2  0.2318103753533321  \\
            0.3  0.21018408508893235  \\
            0.4  0.1832354428227024  \\
            0.5  0.14986073645350723  \\
            0.6  0.10875095889744382  \\
            0.7  0.060181447788077186  \\
            0.8  0.007122683911673172  \\
            0.9  -0.041683639901312804  \\
            1.0  -0.06306346323658707  \\
            1.1  -0.03694572891288634  \\
            1.2  0.04702474834750377  \\
            1.3  0.1520432907401667  \\
            1.4  0.24106748833924013  \\
            1.5  0.31066554847660205  \\
            1.6  0.3808139714624405  \\
            1.7  0.4667810851302487  \\
            1.8  0.5752283499802369  \\
            1.9  0.7098168020190556  \\
            2.0  0.8739093128181401  \\
        }
        ;
    \addlegendentry {biaxial}
    \addplot[mark={none}, dashed, forget plot, blue!60!gray]
        table[row sep={\\}]
        {
            \\
            0.1  0.24912336859803946  \\
            0.2  0.24106748833924035  \\
            0.3  0.22695516827912304  \\
            0.4  0.20556263696261567  \\
            0.5  0.1749304672675303  \\
            0.6  0.13265734370488302  \\
            0.7  0.07749968682388508  \\
            0.8  0.013351747421264715  \\
            0.9  -0.0416836399013128  \\
            1.0  -0.06306346323658707  \\
            1.1  -0.03694572891288633  \\
            1.2  0.047024748347504086  \\
            1.3  0.1524976039344842  \\
            1.4  0.24106748833924035  \\
            1.5  0.31066554847660227  \\
            1.6  0.3808139714624405  \\
            1.7  0.4667810851302487  \\
            1.8  0.5752283499802369  \\
            1.9  0.7098168020190565  \\
            2.0  0.8739093128181414  \\
        }
        ;
    \addplot[mark={none}, thick, green!60!gray]
        table[row sep={\\}]
        {
            \\
            0.1  nan  \\
            0.2  0.30949512540966073  \\
            0.3  0.29032896992156776  \\
            0.4  0.2651741572216219  \\
            0.5  0.23192905043885728  \\
            0.6  0.18729429626589125  \\
            0.7  0.1278186871789492  \\
            0.8  0.0539487681131077  \\
            0.9  -0.025648772399857148  \\
            1.0  -0.06306346323658707  \\
            1.1  -0.01735742817011088  \\
            1.2  0.10857055531394623  \\
            1.3  0.22776522045565267  \\
            1.4  0.315079832322766  \\
            1.5  0.40369733378880085  \\
            1.6  0.5183128944100832  \\
            1.7  0.6677473937843139  \\
            1.8  0.8574693127848043  \\
            1.9  1.0929974378637906  \\
            2.0  1.3801593128638197  \\
        }
        ;
    \addlegendentry {triaxial}
    \addplot[mark={none}, dashed, forget plot, green!60!gray]
        table[row sep={\\}]
        {
            \\
            0.1  0.323839468227417  \\
            0.2  0.31507983232276654  \\
            0.3  0.3003873260524275  \\
            0.4  0.27919687325130566  \\
            0.5  0.249691616690719  \\
            0.6  0.20768013544692554  \\
            0.7  0.14651801896449335  \\
            0.8  0.06276868100856342  \\
            0.9  -0.025648772399857134  \\
            1.0  -0.06306346323658707  \\
            1.1  -0.017357428170110878  \\
            1.2  0.10857055531394666  \\
            1.3  0.227765220455653  \\
            1.4  0.3150798323227663  \\
            1.5  0.4036973337888012  \\
            1.6  0.5183128944100832  \\
            1.7  0.6677473937843139  \\
            1.8  0.8574693127848043  \\
            1.9  1.0929974378637919  \\
            2.0  1.380159312863821  \\
        }
        ;
\end{axis}
\end{tikzpicture}}
    \end{subfigure}
	\caption{Uni-, bi-, triaxial tests based on the undamaged NH (left) and STVK (right) strain energy densities \(\psi^{0}\). The value \(s\) refers to the characterization of the rank-one (uniaxial), rank two (biaxial) and rank three (triaxial) deformation gradient paths as parameterized in \eqref{eq:rankdline}. The dashed lines describe the unrelaxed energy density \(W\) while the solid lines represent the computed rank-one hulls. 
	}
    \label{fig:unibitriaxial}
\end{figure}

We consider a generalization of the boundary value problem of the previous section to three spatial dimensions and show that the proposed convexification procedure is also applicable in this higher, nine-dimensional setting.
However, this has currently been limited to a single increment in a single material point and, thus, corresponds to a single determination of a three-dimensional microstructure.
Again, the off-diagonal entries, the six shear components, have been represented in the mesh \(\mathcal{N}\) by a very coarse discretization.
All parameters related to energy densities and convexification are listed in Table \ref{tab:3Dconvexification}.
\begin{table}
    \smaller
    \begin{tabular}{|c|c|c|c|c|c|c|c|c|c|c|c|}
        \hline
        $\psi^0$ & $\lambda$ & $\mu$ & $D_0$ & $D_\infty$ & $\beta_k$ & $\delta_{ij}$ & 
        $F_{ii}^{\text{min}}$ &
        $F_{ii}^{\text{max}}$ &
        $F_{ij}^{\text{min}} \, \forall i\neq j$ & 
        $F_{ij}^{\text{max}} \, \forall i\neq j$ &
        $|\mathcal{N}|$\\
        \hline
        NH & 0.5 & 1.0 & 0.3 & 0.9 & 0.0625 & 0.15 & 1.0 & 3.4 & -0.15 & 0.15 & 3581577\\
        \hline
        STVK & 0.5 & 1.0 & 0.3 & 0.9 & 0.07 & 0.1 & 0.1 & 2.0 & -0.1 & 0.1 & 5832000\\
        \hline
    \end{tabular}
    \caption{Material and convexification parameters for the $3 \times 3=9$-dimensional convexification for the uni-, bi-, and triaxial deformation configurations.}
    \label{tab:3Dconvexification}
\end{table}
Figure \ref{fig:unibitriaxial} shows slices of the energy densities that correspond to paths of uni-, bi-, and triaxial boundary value problems. These paths are characterized as follows. Let \(\boldsymbol{I}_d \in \R^{d \times d}\) denote the identity matrix. The plotted rank-\(d\) lines are described by
\begin{align} \label{eq:rankdline}
	\boldsymbol{F}_s = \boldsymbol{I}_d + s \, \boldsymbol{I}_d,
\end{align}
for varying \(s\) and \(d = 1, 2, 3\). 
The unrelaxed energy density is dashed and the resulting lines after the higher-dimensional rank-one convexification (in two and three dimensions) and the one-dimensional convexification (in one spatial dimension) are indicated by the red, blue and green line, respectively.
The qualitative behavior is similar, however the quantitative behavior is different due to the higher energy values of the deformations in higher spatial dimensions originating directly from the analytical formulation of \(W\).
This also explains the smaller non-convex regime for the unrelaxed densities in higher spatial dimensions.
It is also worth mentioning that the uniaxial/one-dimensional case results in a convex function, since in this case the underlying path is a rank-one direction.
The higher-dimensional hulls still show non-convexity along the rank-\(d\) paths.
Altogether, this illustrates that the hull obtained in the three-dimensional case reflects the important properties, and even with the very coarse discretization of the nine-dimensional space, the convexification results in a reasonable approximation.

\subsection{Microstructure evolution} \label{sec:examples:microstructure}
\begin{figure}
	\hfill
	\begin{subfigure}[b]{0.49\textwidth}
		\centering
		\ifthenelse{\boolean{professormode}}
		{\includegraphics{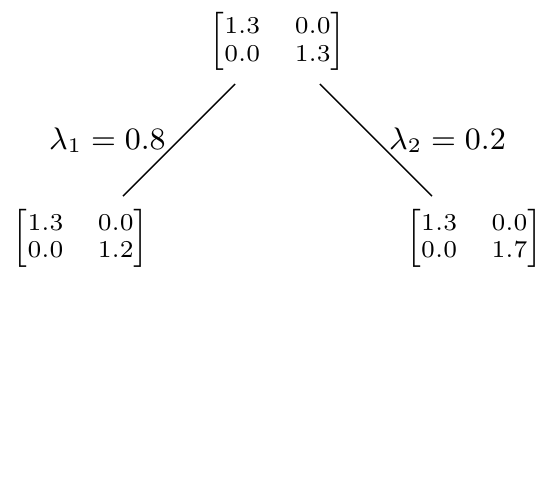}}
		{\tikzsetfigurename{tree1}
\begin{tikzpicture}
	\pgfmathsetmacro\firstsize{2.0}
	\pgfmathsetmacro\secondsize{2.0}
	\pgfmathsetmacro\thirdsize{1.5} 
	\pgfmathsetmacro\verticalsize{2.0}
	\node (F) at (0,0) {\tiny 
		$\begin{bmatrix} 1.3 & 0.0 \\ 0.0 & 1.3 \end{bmatrix}$};
	\node (F1) at (-\firstsize,-\verticalsize) {\tiny
		$\begin{bmatrix} 1.3 & 0.0 \\ 0.0 & 1.2 \end{bmatrix}$};
	\node (F2) at (\firstsize,-\verticalsize) {\tiny
		$\begin{bmatrix} 1.3 & 0.0 \\ 0.0 & 1.7 \end{bmatrix}$};
	\draw (F) -- (F1) node[midway,left] {\footnotesize $\lambda_1=0.8$};
	\draw (F) -- (F2) node[midway,right] {\footnotesize $\lambda_2=0.2$};
	\phantom{
		\node (F12) at (-\firstsize+\secondsize,-2*\verticalsize) {\normalsize
			$\begin{bmatrix} 1.7 & 0.0 \\ 0.0 & 1.2 \end{bmatrix}$};
		\draw (F1) -- (F12) node[midway,right] {\normalsize $\lambda_{12}=0.2$};
		\node (F21) at (\firstsize-\secondsize,-2*\verticalsize) {\normalsize
			$\begin{bmatrix} 1.5 & 0.0 \\ 0.0 & 1.5 \end{bmatrix}$};
		\draw (F2) -- (F21) node[midway,left] {\normalsize $\lambda_{21}=0.33$};
	}
\end{tikzpicture}

%
%
}
		\caption{\(k=1\)}
		\label{fig:tree1}
	\end{subfigure}
	\hfill
	\begin{subfigure}[b]{0.49\textwidth}
		\centering
		\ifthenelse{\boolean{professormode}}
		{\includegraphics{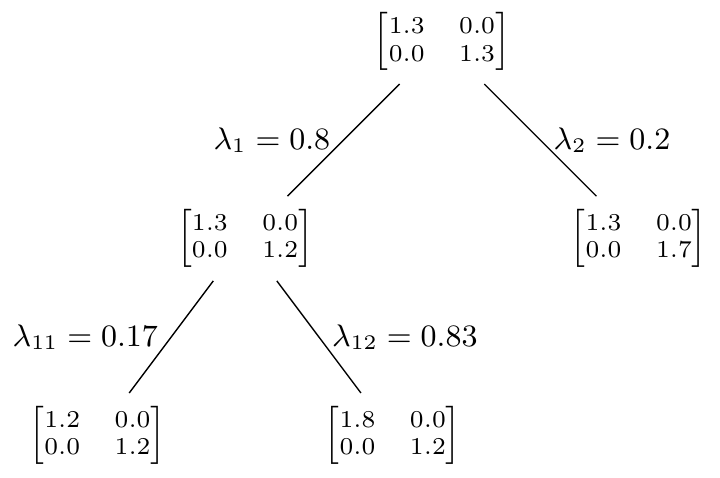}}
		{\tikzsetfigurename{tree2}
\begin{tikzpicture}
	\pgfmathsetmacro\firstsize{2.0}
	\pgfmathsetmacro\secondsize{1.5}
	\pgfmathsetmacro\verticalsize{2.0}
	\node (F) at (0,0) {\tiny 
		$\begin{bmatrix} 1.3 & 0.0 \\ 0.0 & 1.3 \end{bmatrix}$};
	\node (F1) at (-\firstsize,-\verticalsize) {\tiny
		$\begin{bmatrix} 1.3 & 0.0 \\ 0.0 & 1.2 \end{bmatrix}$};
	\node (F2) at (\firstsize,-\verticalsize) {\tiny
		$\begin{bmatrix} 1.3 & 0.0 \\ 0.0 & 1.7 \end{bmatrix}$};
	\draw (F) -- (F1) node[midway,left] {\footnotesize $\lambda_{1}=0.8$};
	\draw (F) -- (F2) node[midway,right] {\footnotesize $\lambda_{2}=0.2$};
	\node (F11) at (-\firstsize-\secondsize,-2*\verticalsize) {\tiny
		$\begin{bmatrix} 1.2 & 0.0 \\ 0.0 & 1.2 \end{bmatrix}$};
	\node (F12) at (-\firstsize+\secondsize,-2*\verticalsize) {\tiny
		$\begin{bmatrix} 1.8 & 0.0 \\ 0.0 & 1.2 \end{bmatrix}$};
	\draw (F1) -- (F11) node[midway,left] {\footnotesize $\lambda_{11}=0.17$};
	\draw (F1) -- (F12) node[midway,right] {\footnotesize $\lambda_{12}=0.83$};
\end{tikzpicture}

%
}
		\caption{\(k=2\)}
		\label{fig:tree2}
	\end{subfigure}
	\hfill \\
	\vspace{3ex}
	\hfill
	\begin{subfigure}[b]{\textwidth}
		\centering
		\ifthenelse{\boolean{professormode}}
		{\includegraphics{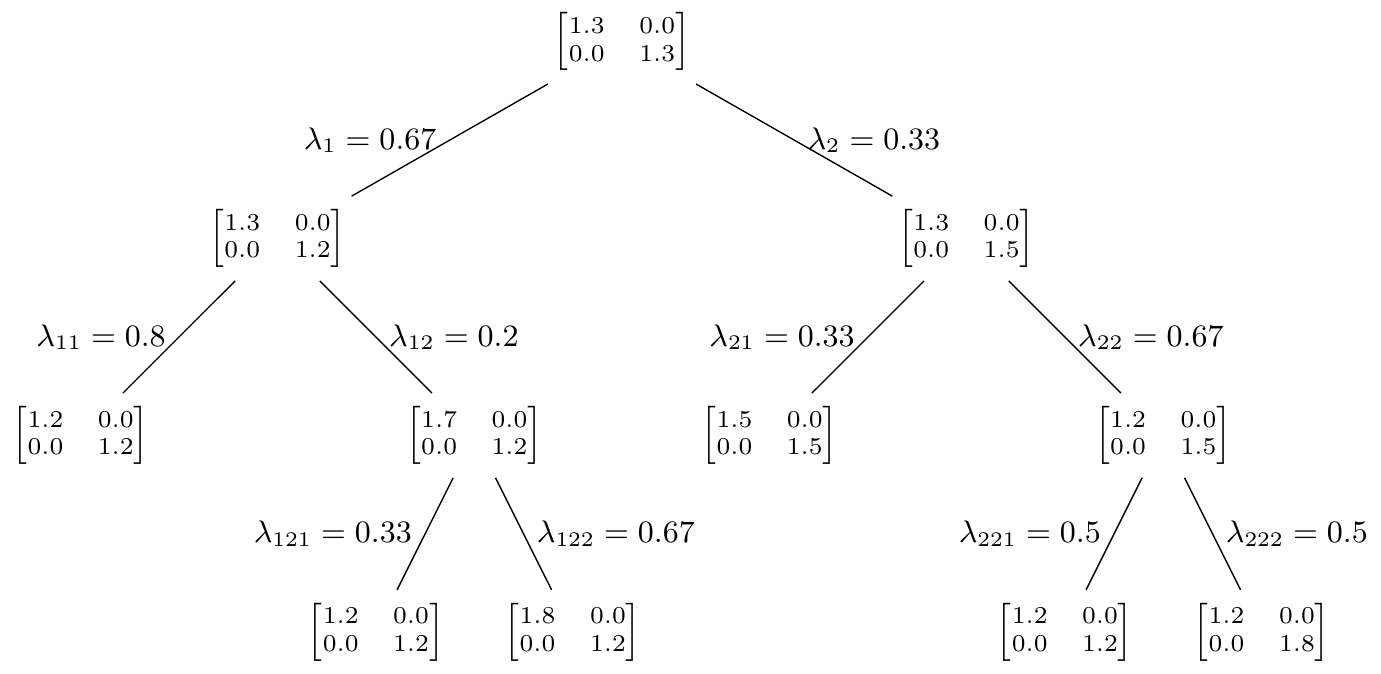}}
		{\tikzsetfigurename{tree3}
\begin{tikzpicture}
	\pgfmathsetmacro\firstsize{3.5}
	\pgfmathsetmacro\secondsize{2.0}
	\pgfmathsetmacro\thirdsize{1.0}
	\pgfmathsetmacro\verticalsize{2.0}
	\node (F) at (0,0) {\tiny 
		$\begin{bmatrix} 1.3 & 0.0 \\ 0.0 & 1.3 \end{bmatrix}$};
	\node (F1) at (-\firstsize,-\verticalsize) {\tiny
		$\begin{bmatrix} 1.3 & 0.0 \\ 0.0 & 1.2 \end{bmatrix}$};
	\node (F2) at (\firstsize,-\verticalsize) {\tiny
		$\begin{bmatrix} 1.3 & 0.0 \\ 0.0 & 1.5 \end{bmatrix}$};
	\draw (F) -- (F1) node[midway,left] {\footnotesize $\lambda_{1}=0.67$};
	\draw (F) -- (F2) node[midway,right] {\footnotesize $\lambda_{2}=0.33$};
	\node (F11) at (-\firstsize-\secondsize,-2*\verticalsize) {\tiny
		$\begin{bmatrix} 1.2 & 0.0 \\ 0.0 & 1.2 \end{bmatrix}$};
	\node (F12) at (-\firstsize+\secondsize,-2*\verticalsize) {\tiny
		$\begin{bmatrix} 1.7 & 0.0 \\ 0.0 & 1.2 \end{bmatrix}$};
	\draw (F1) -- (F11) node[midway,left] {\footnotesize $\lambda_{11}=0.8$};
	\draw (F1) -- (F12) node[midway,right] {\footnotesize $\lambda_{12}=0.2$};
	\node (F21) at (\firstsize-\secondsize,-2*\verticalsize) {\tiny
		$\begin{bmatrix} 1.5 & 0.0 \\ 0.0 & 1.5 \end{bmatrix}$};
	\node (F22) at (\firstsize+\secondsize,-2*\verticalsize) {\tiny
		$\begin{bmatrix} 1.2 & 0.0 \\ 0.0 & 1.5 \end{bmatrix}$};
	\draw (F2) -- (F21) node[midway,left] {\footnotesize $\lambda_{21}=0.33$};
	\draw (F2) -- (F22) node[midway,right] {\footnotesize $\lambda_{22}=0.67$};
	%
	%
	\node (F121) at (-\firstsize+\secondsize-\thirdsize,-3*\verticalsize) {\tiny
		$\begin{bmatrix} 1.2 & 0.0 \\ 0.0 & 1.2 \end{bmatrix}$};
	\node (F122) at (-\firstsize+\secondsize+\thirdsize,-3*\verticalsize) {\tiny
		$\begin{bmatrix} 1.8 & 0.0 \\ 0.0 & 1.2 \end{bmatrix}$};
	\draw (F12) -- (F121) node[midway,left] {\footnotesize $\lambda_{121}=0.33$};
	\draw (F12) -- (F122) node[midway,right] {\footnotesize $\lambda_{122}=0.67$};
	%
	%
	\node (F221) at (+\firstsize+\secondsize-\thirdsize,-3*\verticalsize) {\tiny
		$\begin{bmatrix} 1.2 & 0.0 \\ 0.0 & 1.2 \end{bmatrix}$};
	\node (F222) at (+\firstsize+\secondsize+\thirdsize,-3*\verticalsize) {\tiny
		$\begin{bmatrix} 1.2 & 0.0 \\ 0.0 & 1.8 \end{bmatrix}$};
	\draw (F22) -- (F221) node[midway,left] {\footnotesize $\lambda_{221}=0.5$};
	\draw (F22) -- (F222) node[midway,right] {\footnotesize $\lambda_{222}=0.5$};
\end{tikzpicture}


}
		\caption{\(k=3\)}
		\label{fig:tree3}
	\end{subfigure}
	\hfill
	\caption{Rank-one trees of the deformation gradient \({\boldsymbol{F} = \diag[1.3, 1.3]}\) for the lamination iterations \(k=1\) (\subref{fig:tree1}) and \(k=2\) (\subref{fig:tree2}) and \(k = 3\) (\subref{fig:tree3}). 
	Shown is a special example which contains only branching subject to lamination, due to special setting where the interpolation delivers clearly one favoured nearest point.
	}
	\label{fig:trees}
\end{figure}
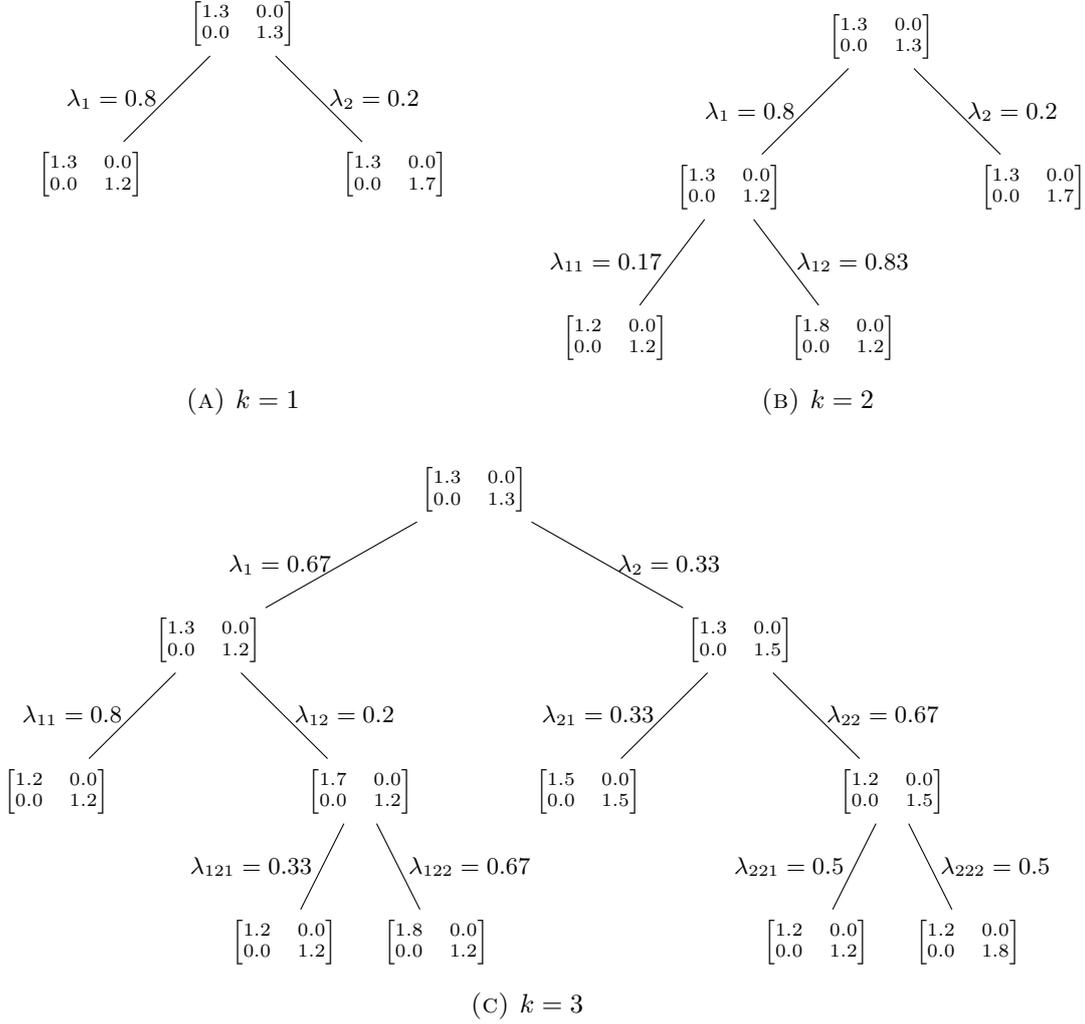

The rank-one trees obtained in the numerical construction of the rank-one convex envelope can be utilized to give a structural interpretation on the microscale.
We consider an energy density \(W\) based on the biaxial test of Section \ref{sec:examples:biaxial} in two spatial dimensions. 
For fixed \(\boldsymbol{F}\), Figure \ref{fig:trees} shows the computed rank-one trees for the first lamination orders \(k=1,2,3\), which correspond to the first three iterations of the convexification algorithm. 
Notice, that the tree in iteration \(k=3\) contains major changes in comparison to the previous iteration.
This is in line with the fact that the new rank-one tree is not necessarily an extension of the previous iterate tree.
This is due to the fact that a different minimum can be found when higher lamination orders are allowed.

The illustrations in Figure \ref{fig:trees} have been simplified and only contain branching induced by lamination and not due to interpolation. The trees can be utilized to show that the leaves form an \(\mathcal{H}_M\) sequence, where \(M\) equals the number of leaves, i.e. \(M = 2\) for \(k = 2\), \(M = 3\) for \(k = 2\), \(M = 6\) for \(k = 3\).
In order to show that these trees visualize \(\mathcal{H}_M\) sequences, we have to come up with suitable \(\xi\)'s as introduced in the definition in Section~\ref{sec:RankOne}.
These values can be interpreted as volume fractions of the leaves in the microstructure and can be computed by the product of the convex coefficients (the \(\lambda\)'s) along the paths, e.g.~in the case \(k = 2\) the volume fraction associated to the leave node \(\diag[1.2, 1.2]\) is given by \(\xi = \lambda_{1} \cdot \lambda_{11} = 0.136\). 
The volume fractions associated to the other two leave nodes are given by \(\lambda_{1} \cdot \lambda_{12} = 0.664\) and \(\lambda_2 = 0.2\), respectively.

\begin{figure}
	\includegraphics[width=1.0\textwidth]{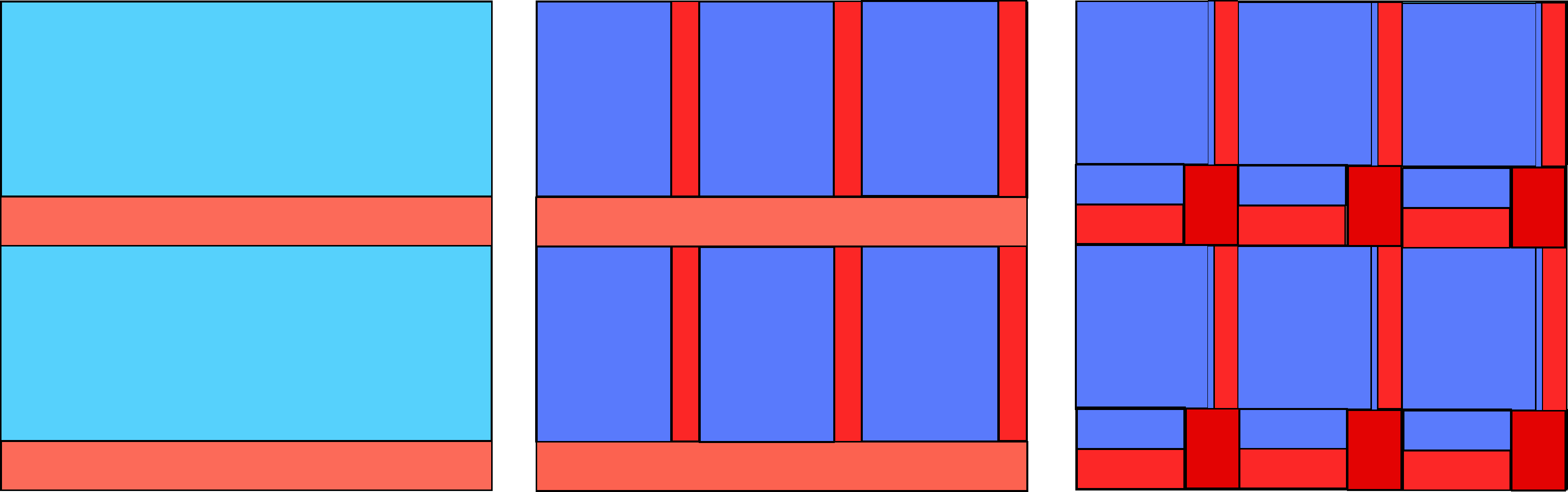}
	\caption{Microstructures associated to the trees in Figure \ref{fig:trees} for $k = 1$ (left), \(k = 2\) (mid), \(k = 3\) (right) and the fixed deformation gradient \(\boldsymbol{F} = \diag[1.3, 1.3]\).
	Note that the microstructures are described by gradient Young measures and, thus, do not encode a length scale or frequency of the phase oscillations, respectively, but instead only the amplitude and volume fraction of the deformation gradients.
	Weakly damaged phases resulting from small deformations are coloured in blue, stronger damaged phases resulting from large deformations are coloured red.
	}
	\label{fig:microstructure}
\end{figure}

We now utilize the information encoded in the tree structure for a microstructural interpretation.
For each of the trees in Figure \ref{fig:trees}, the associated microstructure is displayed in Figure \ref{fig:microstructure}. 
Each computed laminate is illustrated as follows.
The rank-one direction \(\boldsymbol{R}\), which is the minimizing argument of the minimization problem \eqref{eq:minglobiter} and connects \(\boldsymbol{F}^+\) and \(\boldsymbol{F}^-\), characterizes the branching of the two (the weakly and the strongly damaged) phases. 
The normal vector of each lamination direction is given by the right singular vector corresponding to the only nonzero singular value of the rank-one direction matrix that is associated to the branching. 

For example, the right singular vector corresponding to the non-zero singular value of the rank-one direction matrix \(\diag[0, 1]\) associated to the only splitting in \ref{fig:tree1} is given by \([0, 1]^T\).
The normal vector of this laminate is hence pointing in the \(y\)-direction.

The number of phase oscillations/phase changes in a single laminate can not be determined and is therefore only an exemplary realization; however, the volume fraction is given by the convex coefficients \(\lambda\) and \(1 - \lambda\) associated to the branching. 
Weakly damaged phases (small deformation) are coloured in blue, stronger damaged phases (large deformation) are coloured in red.

Each global iteration \(k\) of the procedure provides the possibility of branching in two phases. 
For the microstructure in the case \(k=2\) (middle of Figure \ref{fig:microstructure}), only the weaker phase splits into a new laminate. 
This time, the normal vector obtained from the non-zero right singularvector to the direction \(\diag[1,0]\) is perpendicular to the  first lamination.
The associated convex coefficients to this branching now influence the illustration in the weaker phase in terms of proportion but again the frequency (three phase changes) is only taken for illustration purposes.

The microstructure for the case \(k = 3\) is conceptually different again, as it was the case in Figure \ref{fig:trees} for the tree representation. 
Here, the first laminate differs in terms of proportion as well as intensity of the phases. 
Each of those first order laminate phases is then refined by one branching, and a second branching in one of the phases. 
Again the orientation and relative fraction of those phases can be inferred by the information given in the tree \ref{fig:tree3}.

Note that the volume fractions \(\xi\) of the leave nodes in each tree can be found as area ratio of the corresponding colour in the illustrated cell in \ref{fig:microstructure}.
Since the volume fractions of all leave nodes add up to one (cf. Section \ref{sec:RankOne}) the mean valued colour of the three cells should match and be associated to the fixed deformation gradient \(\boldsymbol{F}\).

Note that for general cases, e.g. a more complex deformation gradient, the laminates can be orientated diagonally and the resulting microstructures could be more complex.

%

\section{Conclusion}\label{sec:conclusion}
In this paper, we showed that the numerical multidimensional relaxation via rank-one convexification yields mesh independent solutions for two element perturbation tests.
For this purpose, a concurrent relaxation scheme was used, which was embedded into finite element simulations.
To this end, efficient parallelized schemes were presented as well as reliable algorithms for the reconstruction of a continuous derivative, which is needed by the finite element solver.
In addition to that, some strain softening was observed, which resulted from the non-convexity along paths with rank more than one.
Along rank-one paths, the function is convex and thus the stress responses from \cite{BalOrt:2012:riv,SchBal:2016:riv} are obtained.
Further, we showed the necessity of high-order laminates, since only laminates of order five or higher show a converged stress response.
The extension of the algorithm by \cite{Bar:2004:lca,Bar:2015:nmn} with the reduced rank-one direction set of \cite{Dol:1999:ncr, DolWal:2000:ena} was necessary for the concurrency of the relaxation.
The relaxation process allowed interpretation in terms of microstructural damage evolution.
However, due to the computational costs, the presented work is still limited in terms of the applicable boundary value problems.
Future work needs to reduce algorithmic complexity further, by e.g. adaptivity in mesh and direction discretization and utilization of known properties of the energy density.
For the former, pioneering work has been recently presented in \cite{ConDol:2018:ara}, where the given macroscopic deformation gradient is the driving refinement criterion.
Due to the presented tree decomposition, a different adaptive criterion is needed, which resolves around the laminate supporting points $\boldsymbol{F}^+$ and $\boldsymbol{F}^-$.

Despite the model's capabilities to describe strain softening along paths with rank higher than one, the softening response is rather untypical for brittle materials which exhibit large deformations.
Therefore, a generalization of the evolution of microstructure in the sense of \cite{KohBal:2023:emr} to higher spatial dimensions is needed, such that a realistic strain softening behavior can be captured.


\medskip
\noindent
{\bfseries Conflicts of Interest}

\noindent
The authors declare that they have no conflicts of interest.

\bibliographystyle{alpha}
\bibliography{ref_peterseim,ref_balzani}


\end{document}